\documentclass[iop]{emulateapj}
\usepackage{apjfonts}

\input epsf.tex

\begin{document}
\title{RXJ0848.6+4453: The Evolution of Galaxy Sizes and Stellar Populations in a $z=1.27$ Cluster}
\author{Inger J{\o}rgensen$^1$, Kristin Chiboucas$^1$, 
Sune Toft$^2$, 
Marcel Bergmann$^3$, 
Andrew Zirm$^2$, 
Ricardo P.\ Schiavon$^{1,4}$, Ruth Gr\"{u}tzbauch$^5$}

\affil{$^1$Gemini Observatory, 670 N.\ A`ohoku Pl., Hilo, HI 96720, USA; 
$^2$Dark Cosmology Centre, Niels Bohr Institute, University of Copenhagen, Juliane Mariesvej 30, DK-2100 Copenhagen, Denmark;
$^3$National Optical Astronomy Observatory, Tucson, AZ 85719, USA;
$^4$Liverpool John Moores University, UK;
$^5$Center for Astronomy and Astrophysics, University of Lisbon, Portugal}

\email{ijorgensen@gemini.edu, kchiboucas@gemini.edu}
\email{sune@dark-cosmology.dk, marcelbergmann@gmail.com}
\email{andrewzirm@gmail.com, R.P.Schiavon@ljmu.ac.uk}
\email{ruthregenborgen@gmail.com}

\submitted{Submitted April 23, 2014. Accepted for publication in Astronomical Journal, August 2, 2014}

\begin{abstract}
RXJ0848.6+4453 (Lynx W) at redshift 1.27 is part of the Lynx Supercluster of galaxies. We present
an analysis of the stellar populations and star formation history for a 
sample of 24 members of the cluster. Our study is based on deep optical spectroscopy
obtained with Gemini North combined with imaging data from {\it Hubble Space Telescope}.
Focusing on the 13 bulge-dominated galaxies for which we can determine central velocity dispersions,
we find that these show a smaller evolution with redshift of sizes and velocity dispersions than reported for field galaxies
and galaxies in poorer clusters.
Our data show that the galaxies in RXJ0848.6+4453 populate the Fundamental Plane (FP) similar to that found for lower redshift clusters. 
The zero point offset for the FP is smaller than expected if the cluster's galaxies are to evolve passively
through the location of the FP we in our previous work established for $z=0.8-0.9$ cluster galaxies and 
then to the present day FP.
The FP zero point for RXJ0848.6+4453 corresponds to an epoch of last star formation at $z_{\rm form}= 1.95^{+0.22}_{-0.15}$.
Further, we find that the spectra of the galaxies in RXJ0848.6+4453 are dominated by young
stellar populations at all galaxy masses and in many cases show emission indicating low level on-going star formation.
The average age of the young stellar populations as estimated from the strength
of the high order Balmer line H$\zeta$ is consistent with a major star formation episode
1-2 Gyr prior, which in turn agrees with  $z_{\rm form}=1.95$.
These galaxies dominated by young stellar populations are distributed throughout the cluster.
We speculate that low level star formation has not yet 
been fully quenched in the center of this cluster maybe because the cluster is significantly
poorer than other clusters previously studied at similar redshifts, which appear to have very little 
on-going star formation in their centers.
The mixture in RXJ0848.6+4453 of passive galaxies with young stellar populations and massive galaxies still experiencing 
some star formation appears similar to the galaxy populations recently identified in two $z\approx 2$ clusters.
\end{abstract}


\section{Introduction}

Recent results regarding stellar populations and sizes of galaxies at 
redshifts above one indicate that the redshift interval $z=1-2$ spans the epoch during which
major changes of galaxy properties took place. Some time during this epoch,
the galaxy sizes change by a factor 3-5 (e.g., Toft et al.\ 2009, 2012; Newman et al.\ 2012; 
Cassata et al.\ 2013; van der Wel et al.\ 2014),
major and minor merging takes place, triggering starbursts that then get quenched
through processes as strangulation and ram-pressure stripping
as the galaxies enter the dense environments of the cluster cores
(Gr\"{u}tzbauch et al.\ 2011; Quadri et al.\ 2012; Koyama et al.\ 2013).
This results in the massive (Mass $> 10^{11}M_{\sun}$)
passively evolving galaxies being in place by
$z \approx 1$, while the lower mass galaxies continue to be added to the passive population as
late as $z=0.5$ (e.g., S\'anchez-Bl\'azquez et al.\ 2009).
Massive clusters of galaxies have recently been found to redshift $z \approx 2$
(e.g., Stanford et al.\ 2012; Gobat et al.\ 2013).
The current studies of these clusters are based primarily on photometry and 
limited redshift information, as very little high signal-to-noise spectroscopy exists of clusters galaxies with $z=1-2$.
However, the results have raised a number of questions fundamental to our understanding of
galaxy evolution as it takes place in dense environments:
\begin{enumerate}
\item
At which epoch does the main size and structure evolution of cluster galaxies take place, and
how is it linked to cluster properties (density, mass, virialization) and galaxy mass?
Results at $z=0.9$ (J\o rgensen \& Chiboucas 2013) and for $z=1.6-2.1$ protoclusters
(Zirm et al.\ 2012; Papovich et al.\ 2012; Strazzullo et al.\ 2013) 
suggest that the size evolution may be accelerated in dense cluster environments. 
However, Newman et al.\ (2014) in a study of a cluster at $z=1.8$ find no difference 
between cluster and field galaxy sizes at this redshift.
\item
When do the first low mass (${\rm Mass} <10^{11}M_{\sun}$) galaxies populate the 
Fundamental Plane (FP) of passive galaxies (Djorgovski \& Davis 1987; J\o rgensen et al.\ 1996) 
as we observe it at $z<1$?  
The prediction from our results for $z<1$ clusters is that the low mass end of the FP is being populated at
$z\approx$1-1.5, while ages from the Balmer lines indicate a much earlier epoch for this process
(J\o rgensen et al.\ 2006, 2007; J\o rgensen \& Chiboucas 2013).
Studies of stellar populations in cluster galaxies at $z>1$ are needed to resolve this issue.
\item
At which epoch and cluster density are the starbursts quenched leading to a large fraction
of post-starburst galaxies, and subsequently to passive galaxies?
Results for protoclusters at $z\approx 2$ show that the main transformation must happen 
at $z=1-2$ and depends on galaxy mass (Quadri et al.\ 2012; Koyama et al.\ 2013)
and possibly also on the cluster environment (Tanaka et al.\ 2013).
\end{enumerate}

\begin{figure*}
\epsfxsize 17.5cm
\epsfbox{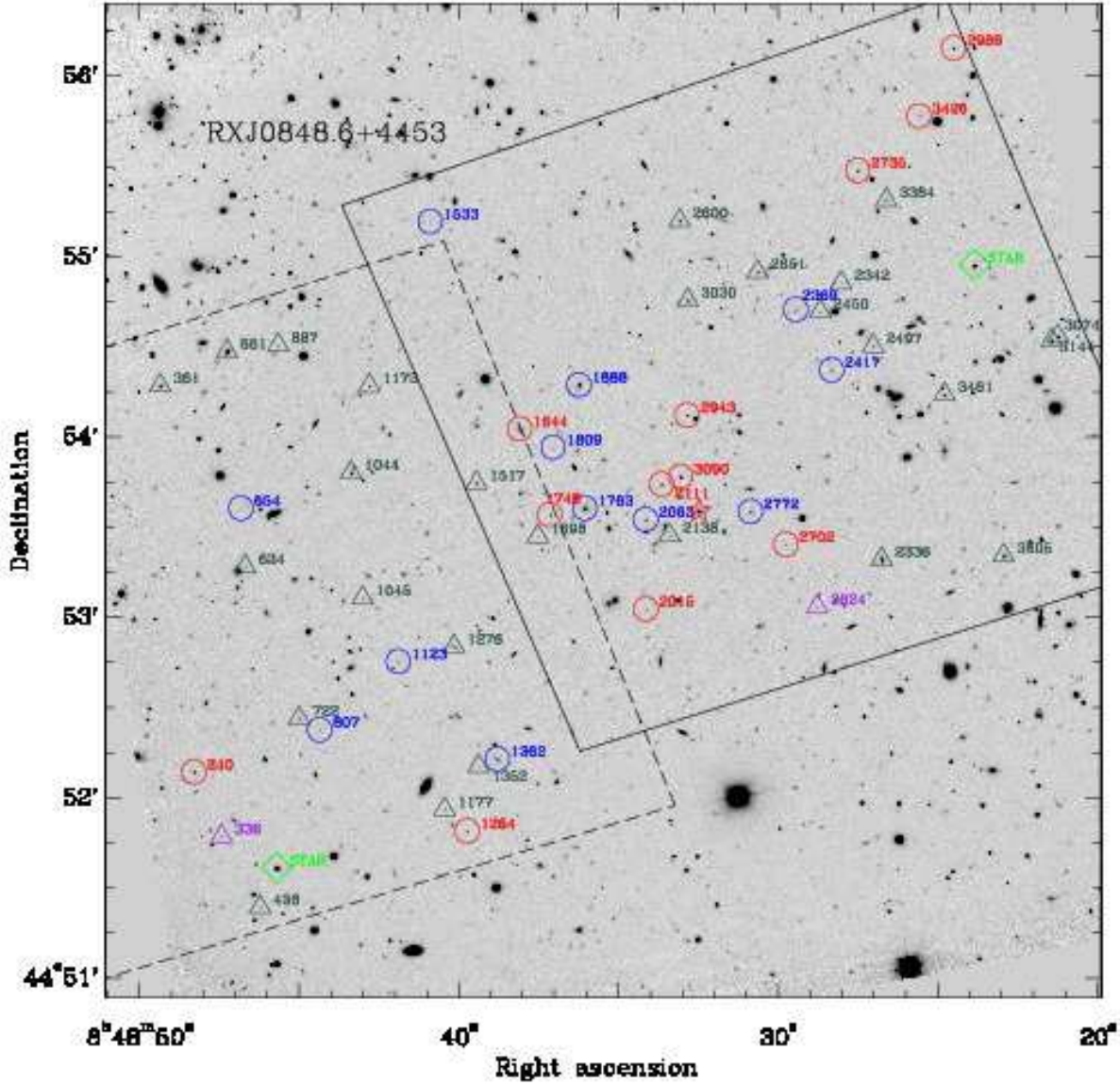}
\caption[]{
GMOS-N $z'$-band image of RXJ0848.6+4453 with our spectroscopic sample marked.
Red circles -- confirmed members with EW[\ion{O}{2}]$\le 5${\AA}. Blue circles -- confirmed members with EW[\ion{O}{2}] $> 5${\AA}.
ID 2063 hosts an AGN, see text.
Dark green triangles -- confirmed non-members. Purple triangles -- targets for which the 
spectra do not allow redshift determination. 
Green diamonds -- blue stars included in the mask to facilitate correction for telluric absorption lines. 
Red star -- the brightest cluster galaxy (BCG), not part of our spectroscopic sample as it has a triple core, see text.
The approximate location of two {\it HST}/ACS fields are marked with black lines.
\label{fig-RXJ0848grey} }
\end{figure*}

Imaging data and deep spectroscopic data of cluster galaxies at $z\approx 1-2$
make it possible to address these questions in detail. With the installation of the 
red sensitive E2V Deep Depletion Charge-Coupled-Devices (E2V DD CCDs) in the
Gemini Multi-Object Spectrograph on Gemini North (GMOS-N) 
it is now possible to obtain such spectra in the rest-frame blue and visible of galaxies up to $z\approx 1.3$.
See Hook et al.\ (2004) for a detailed description of GMOS-N.
For galaxies at higher redshift, such observations are becoming feasible with near-infrared
multi-object spectrographs, e.g., the K-band Multi-Object-Spectrograph (KMOS) on the Very Large Telescope.

We have undertaken a project to obtain such deep spectroscopic observations of rich galaxy
clusters at $z=1-2$. In this paper we present the results from our first observations,
addressing the outlined questions using new high signal-to-noise (S/N) spectra in the 
rest frame 3500-4400 {\AA}ngstrom for individual galaxies in the cluster RXJ0848.6+4453 (Lynx W)
at $z=1.27$. 
We analyze the spectroscopic data together with available {\it Hubble Space Telescope} ({\it HST}) imaging 
of the cluster obtained with the Advanced Camera for Surveys (ACS) in the filters
F775W and F850LP.

The observational data are described in Section \ref{SEC-DATA} and in the Appendix. 
In Section \ref{SEC-CLUSTERZ} we establish the cluster redshift and velocity dispersion 
as well as cluster membership for the observed galaxies.
We also discuss mass of the cluster compared with masses of the other clusters included
in the analysis.
Section \ref{SEC-METHODSAMPLE} gives an overview over the methods and models used in the analysis and 
defines the sub-samples of galaxies, which we refer to throughout the presentation of the results
and the discussion.
Our main results regarding evolution of size and velocity dispersions as well as stellar populations
are described in Section \ref{SEC-RESULTS}.
In Section \ref{SEC-DISCUSSION} we discuss these results in the context of other recent
results for galaxies at $z>1$ as well as simple models for the evolution with redshift. 
The conclusions are summarized in Section \ref{SEC-CONCLUSION}.

Throughout this paper we adopt a $\Lambda$CDM cosmology with 
$\rm H_0 = 70\,km\,s^{-1}\,Mpc^{-1}$, $\Omega_{\rm M}=0.3$, and $\Omega_{\rm \Lambda}=0.7$.

\section{Observational data \label{SEC-DATA}}

\begin{deluxetable*}{lllll}
\tablecaption{GMOS-N Observations\label{tab-GMOSobs} }
\tablewidth{0pt}
\tablehead{
\colhead{Cluster} & \colhead{Program ID} & \colhead{Dates [UT]} & \colhead{Data type} & \colhead{Program type} } \\
\startdata
RXJ0848.6+4445  & GN-2011B-DD-3\tablenotemark{a}  & UT 2011 October 1 to 2011 October 2 & imaging & DD in queue\tablenotemark{c} \\
                & GN-2011B-DD-3\tablenotemark{b}  & UT 2011 December 6 & imaging & DD in queue\tablenotemark{c} \\
                & GN-2011B-DD-5  & UT 2011 November 24 to 2012 January 4 & spectroscopy & DD in queue\tablenotemark{c} \\
                & GN-2013A-Q-65  & UT 2013 March 9 to 2013 May 18 & spectroscopy & queue \\
\enddata
\tablenotetext{a}{Observations obtained with the original E2V CCDs.}
\tablenotetext{b}{Observations obtained with the E2V DD CCDs.}
\tablenotetext{c}{Director's Discretionary time.}
\end{deluxetable*}

\begin{deluxetable*}{lclcc}
\tablecaption{{\it HST}/ACS Imaging Data\label{tab-hstdata} }
\tablewidth{0pc}
\tablehead{
\colhead{Cluster} & \colhead{No. of fields} & \colhead{Filter} & \colhead{Total $t_{exp}$(s)} & \colhead{Program ID} }
\startdata
RXJ0848.6+4453 F1  &  6 & F775W & 7300 & 9919 \\
RXJ0848.6+4453 F2  &  6 & F775W & 7300 & 9919 \\
RXJ0848.6+4443 F1  & 10 & F850LP& 12220 & 9919 \\
RXJ0848.6+4443 F2  & 10 & F850LP& 12220 & 9919 \\
\enddata
\end{deluxetable*}

\begin{deluxetable*}{llrrrrrrr}
\tablecaption{GMOS-N Spectroscopic Data \label{tab-spdata} }
\tablewidth{0pc}
\tablehead{
\colhead{Cluster} & \colhead{Program ID} & \colhead{Exposure time} & \colhead{$N_{\rm exp}$\tablenotemark{a}} & \colhead{FWHM\tablenotemark{b}} & \colhead{$\sigma _{\rm inst}$\tablenotemark{c}} & \colhead{Aperture\tablenotemark{d}} & \colhead{Slit lengths} & \colhead{S/N\tablenotemark{e}} \\ 
\colhead{}        & \colhead{} & \colhead{} & \colhead{} & \colhead{(arcsec)} & \colhead{} & \colhead{(arcsec)} & \colhead{(arcsec)} }
\startdata
RXJ0848.6+4453  & GN-2011B-DD-5 & 39,600 sec & 22  & 0.56 &  \nodata & $1\times 0.7$, 0.53 & 2.75 &   \\
                & GN-2013A-Q-65 & 43,200 sec & 24  & 0.58 &  \nodata & $1\times 0.7$, 0.53 & 2.75 &   \\
                & combined & 82,800 sec & 46  & 0.57 &  3.013\AA, 100 $\rm km\,s^{-1}$ & $1\times 0.7$, 0.53 & 2.75 & 13.3\\
\enddata
\tablenotetext{a}{Number of individual exposures.}
\tablenotetext{b}{Image quality measured as the average FWHM at 8000{\AA} of the blue stars included in the masks.}
\tablenotetext{c}{Median instrumental resolution derived as sigma in Gaussian fits to the sky lines of the stacked 
spectra. The second entry is the equivalent resolution in $\rm km\,s^{-1}$ at 4000{\AA} in the rest frame of the cluster.}  
\tablenotetext{d}{The first entry is the rectangular extraction aperture 
(slit width $\times$ extraction length). The second entry is the radius in an equivalent 
circular aperture, $r_{\rm ap}= 1.025 (\rm {length} \times \rm{width} / \pi)^{1/2}$, cf.\ J\o rgensen et al.\ (1995b).}
\tablenotetext{e}{Median S/N per {\AA}ngstrom for the 24 cluster members, in the rest frame of the cluster. }
\end{deluxetable*}

\begin{figure}
\epsfxsize 8.5cm
\epsfbox{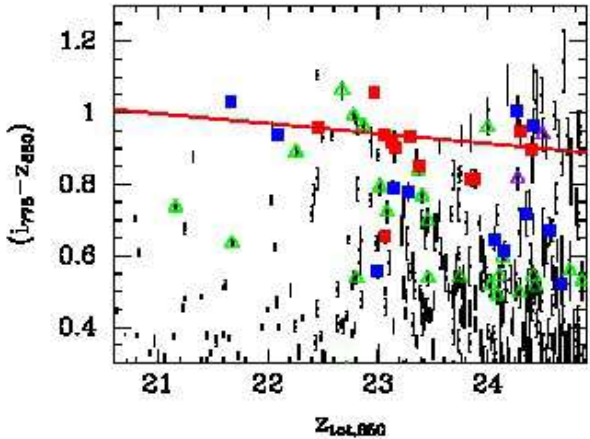}
\caption[]{
Color-magnitude relation of galaxies in the RXJ0848.6+4453 field. 
Red squares -- members of the cluster with EW[\ion{O}{2}]$\le 5${\AA}; blue squares -- members with EW[\ion{O}{2}] $> 5${\AA}.
green open triangles -- confirmed non-members; purple open triangles -- galaxies for which the obtained spectra do not allow 
redshift determination; small black points -- all galaxies in the field.
Red line is the best fit to the 16 member galaxies with $(i_{775}-z_{850})\ge0.8$: 
$(i_{775}-z_{850}) = 0.969 - (0.028 \pm 0.024) ( z_{\rm tot,850} - 22.5)$ with rms=0.073. 
\label{fig-CM} }
\end{figure}

\subsection{Imaging of RXJ0848.6+4453}

Ground-based imaging of RXJ0848.6+4453 was obtained primarily to show the 
performance gain provided by
the replacing the original E2V CCDs in GMOS-N with E2V DD CCDs. This replacement
was done in October 2011.
Imaging of RXJ0848.6+4453 was obtained with the original E2V CCDs in October 2011 and repeated 
with the E2V DD CCDs in November 2011 (Table \ref{tab-GMOSobs}).
For the results presented in this paper we made use of the imaging
for the mask designs for the spectroscopy and to illustrate the spectroscopic sample (Figure \ref{fig-RXJ0848grey}). 
The imaging was done in the $z'$ filter. For the observations with the original E2V CCDs the total
exposure time was 60 min (obtained as twelve 5 min exposures) and the co-added
image had an image quality of FWHM=0.52 arcsec measured from point sources in the field.
For the E2V DD CCDs a total exposure time of 55 min obtained and the resulting image quality was
FWHM=0.51 arcsec.
As {\it HST}/ACS imaging in two filters is available of all galaxies in the spectroscopic sample, no other use is made 
of the ground-based imaging.

Table \ref{tab-hstdata} summarizes the {\it HST}/ACS data for the two fields used in this paper. Data are also available
for a third field, covering Lynx E. However, as none of our spectroscopic sample galaxies
are within that field these data are not used in the present paper. 
We also do not use the shallower imaging obtained for {\it HST} program ID 10496.

The {\it HST}/ACS data were processed as described in Chiboucas et al.\ (2009), 
using the drizzle technique (Fruchter \& Hook 2002).
The images were then processed with SExtractor (Bertin \& Arnouts 1996). Total magnitudes $z_{tot,850}$ 
were derived from the F850LP imaging. Aperture colors $(i_{775}-z_{850})$ were derived within an aperture
with diameter of 0.5 arcsec.
We used GALFIT  (Peng et al.\ 2002) to fit $r^{1/4}$ profiles and S\'{e}rsic (1968) profiles to the galaxies
in the spectroscopic sample and derive effective radii, magnitudes and
surface brightnesses.  This processing was done for the observations in F850LP only.
The effective radii in Table \ref{tab-photRXJ0848HST} are derived from the semi-major and -minor axes as
$r_{\rm e} = (a_{\rm e}\,b_{\rm e})^{1/2}$. The difference between the effective radii derived
from the fits with the  $r^{1/4}$ profiles and S\'{e}rsic profiles as expected correlates with
the S\'{e}rsic index, see details in the Appendix.
The median uncertainty on the S\'{e}rsic index is 0.1, with a few values up to 0.7 for galaxies with
best fit indices larger than 4. As the S\'{e}rsic fits and indices in this paper are 
used primarily for selection of the bulge-dominated galaxies, these uncertainties do not affect our results significantly.

The photometry is calibrated to the AB system using the synthetic zero points for the filters,
25.654 for F775W, 24.862 for F850LP  (Sirianni et al.\ 2005).
Galactic reddening in the direction of the cluster is  E(B-V)=0.024 (Schlafly et al.\ 2011), which
gives, $A_{i_{775}}=0.046$ and $A_{z_{850}}=0.034$.  
The photometric parameters derived using SExtractor and GALFIT are listed in the Appendix, Table \ref{tab-photRXJ0848HST}.

The photometry was calibrated to rest frame B-band.  The calibration was established using
stellar population models from Bruzual \& Charlot (2003) as described in J\o rgensen et al.\ (2005).
The calibration is given in the Appendix.

\begin{deluxetable}{lr}
\tablecaption{Gemini North Instrumentation for Spectroscopic Observations \label{tab-inst} }
\tablewidth{230pt}
\tablehead{}
\startdata
Instrument      & GMOS-N       \\
CCDs            & 3 $\times$ E2V DD 2048$\times$4608 \\
r.o.n.\tablenotemark{a}          & (3.17,3.22,3.46) e$^-$       \\
gain\tablenotemark{a}            & (2.31,2.27,2.17) e$^-$/ADU  \\
Pixel scale     & 0.0727arcsec/pixel \\
Field of view   & $5\farcm5\times5\farcm5$ \\
Grating         & R400\_G5305 \\
Spectroscopic filter & OG515\_G0306 \\
Wavelength range\tablenotemark{b} & 5500-10500\AA  \\
\enddata
\tablenotetext{a}{Values for the three detectors in the array.}
\tablenotetext{b}{The exact wavelength range varies from slitlet to slitlet.}
\end{deluxetable}

\subsection{Spectroscopy of RXJ0848.6+4453\label{SEC-SPECSEL}}

The spectroscopic observations were obtained in multi-object spectroscopic (MOS) mode with GMOS-N,
see Table \ref{tab-GMOSobs} for the dates of the observations.
The sample selection is based on the photometry from the {\it HST}/ACS imaging.
Figure \ref{fig-CM} shows the color-magnitude (CM) relation for the field.
Galaxies were selected to maximize the coverage along the red sequence from the 
brightest cluster galaxy (BCG) to 
$z_{\rm tot,850} \approx 24.5$ mag.
The triple merger (red star on Fig.\ \ref{fig-RXJ0848grey}),
which is considered the BCG, was not included in the observations. 
This decision was made in part because it would eliminate another known bright 
cluster member from the mask and in part because 
the angular separation of the components is only $\approx 0.5$ arcsec making it
very difficult to separate them in ground-based spectroscopy obtained in natural seeing.
Priority was given to galaxies within 0.1 mag of the CM relation in $(i_{775}-z_{850})$ versus $z_{\rm tot,850}$.
Additional space in the mask was filled with galaxies with $(i_{775}-z_{850})>0.5$ and $z_{\rm tot,850}$ 
in the interval from 21 mag to 25 mag. Some of these turned out to be blue cluster members.
The spectroscopic sample is marked on Figures \ref{fig-RXJ0848grey} and \ref{fig-CM}.
Two blue stars were included in the mask in order to obtain a good correction
for the telluric absorption lines. The blue stars are also marked on Figure \ref{fig-RXJ0848grey}.

Table \ref{tab-spdata} gives an overview of the obtained spectroscopic observations,
while Table \ref{tab-inst} summarizes the instrument parameters for the observations.
After the initial observations from program GN-2011B-DD-5 non-members identified from those observations
were eliminated in the mask made for program GN-2013A-Q-65 and other potential members were included instead
using the same selection criteria as used for the original mask.

The spectroscopic observations were processed using the methods described in detail in J\o rgensen \& Chiboucas (2013). 
The only exception was the handling of the charge diffusion effect, which for the E2V DD CCDs turns out to have
a spatial dependence in addition to variation with wavelength. The adopted correction is described in the Appendix.

The data processing results in 1-dimensional spectra calibrated to a relative flux scale. 
The spectra were used to derive redshifts, and for those targets with sufficient S/N,
velocity dispersions and absorption line strengths.
The spectroscopic parameters were determined using the same methods as described in J\o rgensen et al.\ (2005).
In particular, the redshifts and velocity dispersions were determined by fitting the spectra with
a mix of three template stars (spectral types K0III, G1V, and B8V) using software made available by Karl Gebhardt
(Gebhardt et al.\ 2000, 2003). The galaxies that were found to be members of the cluster were fit in
the wavelength range 3750--4100{\AA}. 
Table \ref{tab-RXJ0848kin} in the Appendix gives the results from the template fitting.
The velocity dispersions were aperture corrected using the technique from J\o rgensen et al.\ (1995b).
Of the 52 targets observed, 24 turned out to be members of the cluster. Of these 19 have velocity 
dispersion determined. The median S/N of these is 18 per {\AA}ngstrom in the rest frame (see Table \ref{tab-RXJ0848kin}
for the S/N of the individual spectra).
Velocity dispersions were determined for 10 non-members, see Table \ref{tab-RXJ0848kin}.
We assessed the possible systematic errors on the derived velocity dispersions resulting from 
the determination of the instrumental resolution, the adopted telluric correction, and the sky subtraction.
None of these sources cause systematic errors that affect our results significantly, see Appendix for details.

The spectra allow determination of following absorption line indices: 
CN3883, CaHK, D4000, and H$\zeta _{\rm A}$. 
The indices CN3883 and CaHK are defined in Davidge \& Clark (1994). For D4000 we use a shorter red passband
than usually used (Gorgas et al.\ 1999) but calibrate our measurements to be consistent with the conventional 
definition of the index, see Appendix for details. 
For the high order Balmer line index H$\zeta _{\rm A}$ we adopt the definition from Nantais et al.\ (2013).
All measured indices are listed in Table \ref{tab-RXJ0848line} in the Appendix.

For galaxies with detectable  [\ion{O}{2}] emission the strength
of the  emission line was determined both as a (relative) flux and as an 
equivalent width (Table \ref{tab-RXJ0848line}). 
Due to the very weak continuum of some of the emission line galaxies,
the equivalent widths in some cases have very large uncertainties.

The [\ion{O}{2}] emission makes it possible to determine the
star formation rates (SFR), under the assumption that the line is due to star formation only.
We have examined the spectra for the presence of the two neon lines [\ion{Ne}{5}]3426{\AA}
and [\ion{Ne}{3}]3869{\AA}. 
The line [\ion{Ne}{5}]3346{\AA} is at the cluster redshift too close to strong telluric absorption to
be reliably detected.
These are the only strong lines originating from active galactic nuclei (AGNs) that are in the 
covered wavelength region (e.g., Schmidt et al.\ 1998; Mignoli et al.\ 2013).
Only ID 2063 have detectable Ne emission. We proceed to determine the SFR from the [\ion{O}{2}] line
and flag ID 2063 in the relevant figures in the analysis of the data.

\subsection{Data for other clusters \label{SEC-COMPDATA} }

In the analysis we use our data for the $z=0.5-0.9$ clusters published in our previous 
papers (J\o rgensen et al.\ 2005; Chiboucas et al.\ 2009; J\o rgensen \& Chiboucas 2013). 
These papers present ground-based spectroscopy and {\it HST}/ACS imaging of
MS0451.6--0305 ($z=0.54$), RXJ0152.7--1357 ($z=0.83$), and RXJ1226.9+3332 ($z=0.89$).
Further, we use our data for Coma, Perseus and Abell 194, 
(J\o rgensen et al.\ 1995ab, 1999; J\o rgensen \& Chiboucas 2013), as our $z\approx 0$ comparison sample.
Table \ref{tab-redshifts} summarizes the cluster properties for all the clusters.
The samples in all clusters are selected consistently.
Except for MS0451.6--0305, the passbands used for the determination of the effective parameters 
correspond roughly to rest frame B-band, as is the case for the RXJ0848.6+4453 observations presented in 
this paper. MS0451.6--0305 was observed in F814W, which at $z=0.54$ is close to rest frame V-band.
Thus, if the color gradients the MS0451.6--0305 galaxies are similar to those in nearby
early-type galaxies, typically $\Delta (B-V)/\Delta \log r \approx -0.04$ (e.g., Saglia et al.\ 2000),
then the effective radii derived from F814W can be expected to be about 6\% smaller than if derived from a 
passband matching rest frame B-band for the cluster (Sparks \& J\o rgensen 1993).

\begin{deluxetable*}{llrrrrr}
\tablecaption{Cluster Properties \label{tab-redshifts} }
\tablewidth{0pt}
\tablehead{
\colhead{Cluster} & \colhead{Redshift} & \colhead{$\sigma _{\rm cluster}$} & 
\colhead{$L_{500}$} & \colhead{$M_{500}$} & \colhead{$R_{500}$} & \colhead{N$_{\rm member}$} \\
 & & $\rm km~s^{-1}$ & $10^{44} \rm{erg\,s^{-1}}$ & $10^{14}M_{\sun}$ & Mpc \\
\colhead{(1)} & \colhead{(2)} & \colhead{(3)} & \colhead{(4)} & \colhead{(5)} & \colhead{(6)} & \colhead{(7)}
}
\startdata
Perseus = Abell 426\tablenotemark{a} & 0.0179 & $1277_{-78}^{+95}$ &   6.217 & 6.151 & 1.286 &  63 \\
Abell 194\tablenotemark{a,b}         & 0.0180 &  $480_{-38}^{+48}$ &   0.070 & 0.398 & 0.516 &  17 \\
Coma = Abell 1656\tablenotemark{a}   & 0.0231 & $1010_{-44}^{+51}$ &   3.456 & 4.285 & 1.138 & 116 \\
MS0451.6--0305\tablenotemark{c}    & $0.5398\pm 0.0010$ & $1450_{-159}^{+105}$ & 15.352 & 7.134 & 1.118 &  47 \\
RXJ0152.7--1357\tablenotemark{d}   & $0.8350\pm 0.0012$ & $1110_{-174}^{+147}$ &  6.291 & 3.222 & 0.763 &  29 \\
RXJ1226.9+3332\tablenotemark{c}    & $0.8908\pm 0.0011$ & $1298_{-137}^{+122}$ & 11.253 & 4.386 & 0.827 &  55 \\
RXJ0848.6+4443\tablenotemark{e}    & $1.2701\pm 0.0010$ & $733_{-85}^{+84}$ & 1.04   & 1.37 & 0.499 &  24 \\
\enddata
\tablecomments{Col.\ (1) Galaxy cluster; col.\ (2) cluster redshift; col.\ (3) cluster velocity dispersion;
col.\ (4) X-ray luminosity in the 0.1--2.4 keV band within the radius $R_{500}$, from Piffaretti et al.\ (2011), except for RXJ0848.6+4443
for which the data are from Ettori et al.\ (2004); 
col.\ (5) cluster mass derived from X-ray data within the radius $R_{500}$, from Piffaretti et al., except for RXJ0848.6+4443
for which the data are from Ettori et al.;
col.\ (6) radius within which the mean over-density of the cluster is 500 times the critical density at the 
cluster redshift, from Piffaretti et al., except for RXJ0848.6+4443
for which the data are from Ettori et al.;
col.\ (7) Number of member galaxies for which spectroscopy is used in this paper.}
\tablenotetext{a}{Redshift and velocity dispersion from Zabludoff et al.\ (1990).}
\tablenotetext{b}{Abell 194 does not meet the X-ray luminosity selection criteria of the main cluster sample.}
\tablenotetext{c}{Redshifts and velocity dispersions from J\o rgensen \& Chiboucas (2013).}
\tablenotetext{d}{Redshift and velocity dispersion from J\o rgensen et al.\ (2005). The velocity dispersions for the Northern and Southern sub-clusters are $(681 \pm 232)\, \rm {km\,s^{-1}}$ and 
$(866 \pm 266)\, \rm {km\,s^{-1}}$, respectively.}
\tablenotetext{e}{Redshift and velocity dispersion from this paper.}
\end{deluxetable*}

\begin{figure}
\epsfxsize 8.5cm
\epsfbox{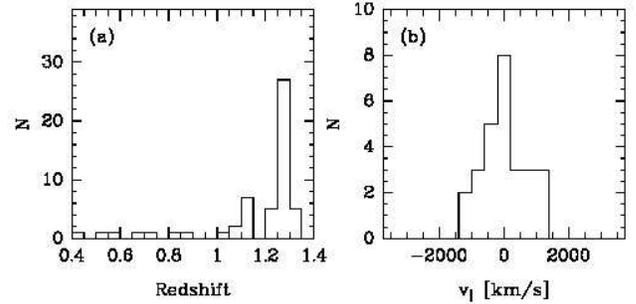}
\caption[]{(a) Redshift distribution of the spectroscopic sample.
(b) Distribution of the radial velocities (in the rest frame of the cluster) 
relative to the cluster redshifts for cluster members, $v_{\|} = c (z - z_{\rm cluster}) / (1+z_{\rm cluster})$.
The radial velocity distributions for the members are not significantly different
from a Gaussian distribution.
\label{fig-zhist} }
\end{figure}

\begin{figure}
\epsfxsize 8.5cm
\epsfbox{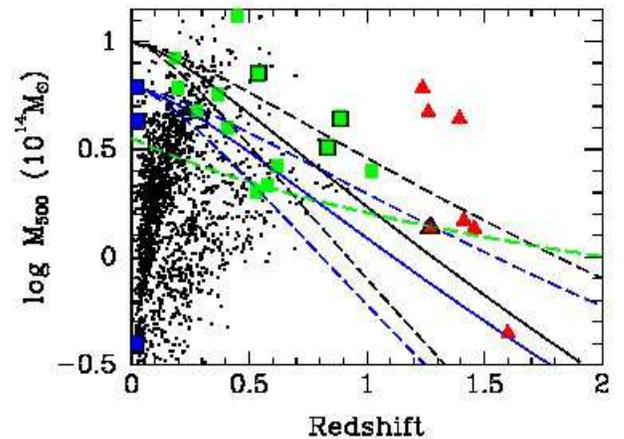}
\caption[]{The cluster masses, $M_{\rm 500}$, based on X-ray data versus the redshifts of the clusters.
Blue -- our low redshift comparison sample (Coma, Perseus, A194); 
green - our main $z=0.2-1$ cluster sample,
MS0451--0305, RXJ0152.7--1357, and RXJ1226.9+3332 shown with black outlines;
$M_{\rm 500}$ from Piffaretti et al.\ (2011).
Red -- our current sample of $z=1.2-1.6$ clusters, RXJ0848.6+4453/Lynx W shown with black outline.
The $z=1.2-1.6$ clusters and references for the cluster masses are: 
RDCS\,J1252.9--2927 (Stott et al.\ 2010),
RXJ0848.6+4453/Lynx W (Ettori et al.\ 2004),
RXJ0848.9+4452/Lynx E (Stott et al.\ 2010),
XMM2235.3--2557 (Rosati et al.\ 2009),
J1438.1+3414 (Brodwin et al.\ 2011),
J2215.0--1738.1 (Stott et al.\ 2010),
J0332--2742 ($M_{\rm 500}$ derived from $L_{\rm 500}$ in Kurk et al.\ 2009, using approximation from
Vikhlinin et al.\ 2009). 
Small black points -- all clusters from Piffaretti et al.\ shown for reference.
Green line -- original X-ray luminosity limit for our $z=0.2-1$ cluster sample (J\o rgensen \& Chiboucas 2013).
Blue and black lines -- mass development of clusters based on numerical simulations
by van den Bosch (2002).
The black lines terminate at Mass=$10^{15} M_{\sun}$ at $z=0$ roughly
matching the highest mass clusters at $z=0.1-0.2$.
The blue lines terminate at Mass=$10^{14.8} M_{\sun}$ at $z=0$
matching the mass of the Perseus cluster.
The dashed lines represent the typical uncertainty in the
mass development represented by the numerical simulations.
\label{fig-M500} }
\end{figure}

\begin{figure}
\epsfxsize 8.5cm
\epsfbox{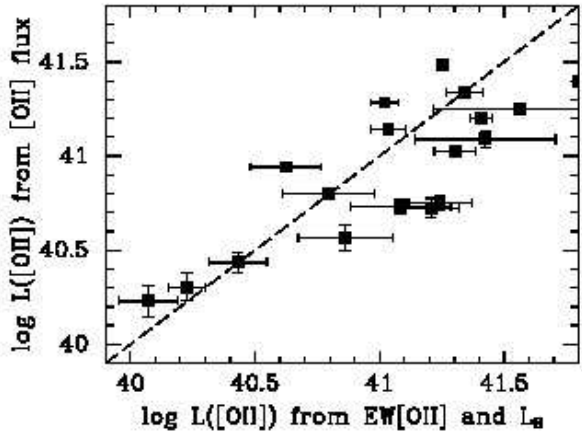}
\caption[]{
The luminosity of the [\ion{O}{2}] emission, $L$([\ion{O}{2}]), derived from EW[\ion{O}{2}] 
and $L_B$ versus the luminosity derived directly from the [\ion{O}{2}] flux.
Dot-dashed line -- one-to-one relation.
Using the [\ion{O}{2}] flux directly results in lower uncertainties on $L$([\ion{O}{2}]). 
The offset of $\Delta \log L$([\ion{O}{2}]) $=0.1$ between the two methods
is of no importance for our analysis. 
The offset due to a combination of the aperture size for the spectroscopy and the 
non-photometric conditions for some of those observations.
\label{fig-SFRdeterm} }
\end{figure}

\begin{deluxetable*}{llrl}
\tablecaption{Predictions from Single Stellar Population Models \label{tab-models} }
\tablewidth{0pc}
\tablehead{
\multicolumn{2}{l}{Relation} & \colhead{rms} & \colhead{Reference} \\
\multicolumn{2}{l}{(1)} & \colhead{(2)} & \colhead{(3)} }
\startdata
$\rm \log M/L_B $ &$= 0.946 \log {\rm age} + 0.333 {\rm [M/H]} - 0.063 $  & 0.022 & Maraston 2005 \\
$\rm \log H\zeta _{A} $ &$= -0.456 \log {\rm age} - 0.462 {\rm [M/H]} + 0.563$  & 0.049 & Maraston \& Str\"{o}mb\"{a}ck 2011 \\
$\rm CN3883 $ &$= 0.183 \log {\rm age} + 0.258 {\rm [M/H]} + 0.097$  & 0.024 & Maraston \& Str\"{o}mb\"{a}ck 2011 \\
$\rm D4000 $ &$= 0.617 \log {\rm age} + 0.807 {\rm [M/H]} + 1.695$  & 0.069 & Maraston \& Str\"{o}mb\"{a}ck 2011 \\
$\rm \log CaHK $ &$= 0.086 \log {\rm age} + 0.079 {\rm [M/H]} + 1.279$  & 0.011 & Maraston \& Str\"{o}mb\"{a}ck 2011 \\
\enddata
\tablecomments{ (1) Relation established from the published model values. 
${\rm [M/H]}\equiv \log Z/Z_\sun$ is the total metallicity relative to solar.
The age is in Gyr. The M/L ratios are stellar M/L ratios in solar units. 
The models were fit for ages from 1 to 15 Gyr and [M/H] from $-0.3$ to 0.3.
The relation for the M/L ratio differs slightly from the relation given in J\o rgensen \& Chiboucas (2013),
as that paper did not include the 1 Gyr old models in the fit.
(2) Scatter of the model values relative to the relation. (3) Reference for the model values. }
\end{deluxetable*}

\section{Cluster redshift, velocity dispersion, substructure, and cluster mass \label{SEC-CLUSTERZ}}

We determined the cluster redshift and velocity dispersion using the bi-weight method (Beers et al.\ 1990).
Figure \ref{fig-zhist} shows the redshift distribution of the sample. We find a redshift of $1.2701\pm0.0010$
and a cluster velocity dispersion of $733_{-85}^{+84} {\rm km\,s^{-1}}$. 
Galaxies with redshifts in the interval $z=1.26-1.28$ are considered members of the cluster, see Table \ref{tab-RXJ0848kin}
in the Appendix for membership information of the individual galaxies.

Using a Kolmogorov-Smirnov test, we tested whether the velocity distribution of the member galaxies deviate 
from a Gaussian.  The probability of the sample being drawn from a Gaussian distribution is 99\%. 
Thus, we conclude that no sub-structure is detectable in the velocity distribution.

Table \ref{tab-redshifts} summarizes the cluster properties for all clusters used in the analysis 
in this paper, including the X-ray luminosities, radii and masses from the literature.
On Figure \ref{fig-M500} we show the cluster masses versus redshifts for these clusters.
For reference the figure also shows our full cluster sample and the catalog of $z<1$ clusters
from Piffaretti et al.\ (2011).
We show sample models for the growth of cluster masses with time, based on results
from van den Bosch (2002). 
These models are in general agreement with newer and more detailed analysis of the results from
the Millennium simulations (Fakhouri et al.\ 2010).
The mass of RXJ0848.9+4453 is significantly lower than the other clusters included in the present analysis.
However, because of the expected growth of cluster masses with time, RXJ0848.9+4453 is a viable
progenitor for clusters of masses similar to Coma and Perseus at $z\approx 0$.

\section{The methods, stellar population models and the final galaxy sample \label{SEC-METHODSAMPLE} }

We characterize the stellar populations in the RXJ0848.6+4453 galaxies by 
(1) establishing the FP, and the relations between masses, sizes and velocity dispersions; and 
(2) analyzing the absorption lines and SFR as function of galaxy velocity dispersion (or mass) and location
within the cluster.
In our analysis of the possible size and velocity dispersion evolution we present results using 
effective radii from both the fits with $r^{1/4}$ profiles and with S\'{e}rsic profiles. 
The results for the FP do not depend which of the two sets of effective parameters are used
as on average the combination $\log r_{\rm e}+\beta \log \langle I \rangle _{\rm e}$ with $\beta=0.66-0.82$ 
entering the FP varies less than 0.01 between the two choices of parameters. Further, since the low 
redshift comparison sample was fit with $r^{1/4}$ profiles we use the effective radii and surface 
brightnesses derived from the fits with $r^{1/4}$ profiles in our discussion of the FP.

\subsection{Fitting the scaling relations}

Our technique for establishing the scaling relations and associated uncertainties on
slopes and zero points is the same as we used in 
J\o rgensen et al.\ (2005) and J\o rgensen \& Chiboucas (2013).
Briefly, we establish the scaling relations using a fitting technique that
minimizes the sum of the absolute residuals perpendicular to the relation,
determines the zero points as the median, and uncertainties on the slopes using
a boot-strap method.
The technique is very robust to the effect of outliers.
The random uncertainties on the zero point differences, $\Delta \gamma$, 
between the intermediate redshift and low redshift samples are derived as
\begin{equation}
\sigma _{\Delta \gamma} = \left ( {\rm rms}_{{\rm low-}z}^2/N_{{\rm low-}z}
  +  {\rm rms}_{{\rm int-}z}^2/N_{{\rm int-}z} \right )^{0.5}
\end{equation}
where subscripts ``low-$z$'' and ``int-$z$'' refer to the low redshift sample
and one of the intermediate redshift clusters, respectively.
In the presentation of the zero point differences we show
only the random uncertainties.
The systematic uncertainties on the zero point differences are expected to
be dominated by the possible inconsistency in the calibration of the 
velocity dispersions, 0.026 in $\log \sigma$ (cf.\ J\o rgensen et al.\ 2005), and may 
be estimated as 0.026 times the coefficient for $\log \sigma$, or
0.052 times the coefficient for $\log {\rm Mass}$.

\begin{deluxetable*}{lllll}
\tablecaption{RXJ0848.6+4453 Sub-Samples\label{tab-sample} }
\tablewidth{0pc}
\tablehead{
\colhead{No.} & \colhead{Criteria} & \colhead{N} & \colhead{IDs} & \colhead{Notes} }
\startdata
1  &  Only redshift \& EW[\ion{O}{2}] from spectra & 5 &
   654 1123 1533 1809 3426 & ID 3426 has no significant [\ion{O}{2}] emission \\
2  &  $n_{\rm ser}<1.5$, $\log \sigma$ meas. & 4 &
   1644 2349 2417 2772 & IDs 1644 2349 2417 have S/N$<$10 \\
3  &  $n_{\rm ser}\ge1.5$, $\log \sigma$ meas., S/N$<$10 & 2 &
   2015 2702 & ID 2015 has log Mass $<$ 10.3 \\
4  &  $n_{\rm ser}\ge1.5$, $\log \sigma$ meas., S/N$\ge$10, EW[\ion{O}{2}]$>$5{\AA}   & 5 &
   807 1362 1763 1888 2063 & log Mass $\ge$ 10.3, ID 2063 hosts an AGN \\
5  &  $n_{\rm ser}\ge1.5$, $\log \sigma$ meas., S/N$\ge$10, EW[\ion{O}{2}]$\le$5{\AA} & 8 &
   240 1264 1748 2111 2735 2943 2989 3090 & log Mass $\ge$ 10.3 \\
\enddata
\end{deluxetable*}

\begin{deluxetable*}{lrrr rrr rrr rrr rrr}
\tablecaption{Scaling Relations \label{tab-relations} }
\tablewidth{0pc}
\tabletypesize{\scriptsize}
\tablehead{
\colhead{Relation} & \multicolumn{3}{c}{Low redshift} & 
  \multicolumn{3}{c}{MS0451.6--0305} &\multicolumn{3}{c}{RXJ0152.7--1357} & \multicolumn{3}{c}{RXJ1226.9+3332}  & \multicolumn{3}{c}{RXJ0848.6+4453} \\
 & \colhead{$\gamma$} & \colhead{$N_{\rm gal}$} & \colhead{rms} 
 & \colhead{$\gamma$} & \colhead{$N_{\rm gal}$} & \colhead{rms} &  \colhead{$\gamma$} & \colhead{$N_{\rm gal}$} & \colhead{rms} 
 & \colhead{$\gamma$} & \colhead{$N_{\rm gal}$} & \colhead{rms} & \colhead{$\gamma$} & \colhead{$N_{\rm gal}$} & \colhead{rms} \\
\colhead{(1)} & \colhead{(2)} & \colhead{(3)} & \colhead{(4)} 
& \colhead{(5)} & \colhead{(6)} & \colhead{(7)} & \colhead{(8)} & \colhead{(9)} & \colhead{(10)} 
& \colhead{(11)} & \colhead{(12)} & \colhead{(13)}
& \colhead{(14)} & \colhead{(15)} & \colhead{(16)} 
}
\startdata
$\log r_e      = (0.57 \pm 0.06) \log {\rm Mass} + \gamma$\tablenotemark{a}  & -5.734 & 105 & 0.16 & -5.701 & 34 & 0.17 & -5.682 & 21 & 0.11 & -5.724 & 28 & 0.19  & -5.806 & 8 & 0.22 \\   
$\log r_e      = (0.57 \pm 0.06) \log {\rm Mass} + \gamma$\tablenotemark{b}  & \nodata & \nodata & \nodata & -5.704 & 34 & 0.17 & -5.715 & 21 & 0.10 & -5.735 & 28 & 0.19  & -5.843 & 8 & 0.27 \\   
$\log r_e      = (0.57 \pm 0.06) \log {\rm Mass} + \gamma$\tablenotemark{c}  & \nodata & \nodata & \nodata & -5.700 & 34 & 0.16 & -5.713 & 21 & 0.10 & -5.725 & 28 & 0.18  & -5.838 & 8 & 0.27 \\   
$\log \sigma   = (0.26 \pm 0.03) \log {\rm Mass} + \gamma$\tablenotemark{a}  & -0.667 & 105 & 0.08 & -0.701 & 34 & 0.08 & -0.716 & 21 & 0.05 & -0.679 & 28 & 0.09  & -0.635 & 8 & 0.10 \\   
$\log \sigma   = (0.26 \pm 0.03) \log {\rm Mass} + \gamma$\tablenotemark{b}  & \nodata & \nodata & \nodata & -0.694 & 34 & 0.08 & -0.684 & 21 & 0.05 & -0.673 & 28 & 0.09  & -0.614 & 8 & 0.14 \\   
$\log \sigma   = (0.26 \pm 0.03) \log {\rm Mass} + \gamma$\tablenotemark{c}  & \nodata & \nodata & \nodata & -0.695 & 34 & 0.08 & -0.686 & 21 & 0.05 & -0.674 & 28 & 0.09  & -0.617 & 8 & 0.14 \\   
\\
$\log \rm{M/L} = (0.24 \pm 0.03) \log {\rm Mass} + \gamma$  & -1.754 & 105 & 0.09 & -2.144 & 34 & 0.15 & -2.289 & 21 & 0.20 & -2.410 & 28 & 0.19  & -2.523 & 8 & 0.18 \\   
$\log \rm{M/L} = (1.07 \pm 0.12) \log \sigma     + \gamma$  & -1.560 & 105 & 0.11 & -1.974 & 34 & 0.13 & -2.067 & 21 & 0.18 & -2.197 & 28 & 0.17  & -2.374 & 8 & 0.20 \\   
$\log \rm{H}\zeta _{\rm A} = (-0.76 \pm 0.29) \log \sigma + \gamma$  & 1.762 & 45 & 0.25 & 1.877 & 28 & 0.16 & 1.956 & 21 & 0.22 & 2.037 & 17 & 0.11  & 1.211 & 6 & 0.27 \\   
$\rm{CN3883}    = (0.29 \pm 0.06) \log \sigma    + \gamma$  & -0.410 & 65 & 0.05 & -0.410 & 31 & 0.03 & -0.396 & 21 & 0.05 & -0.400 & 23 & 0.04  & -0.512 & 6 & 0.06 \\   
$\log \rm{CaHK} = (0.14 \pm 0.04) \log \sigma    + \gamma$  &  0.997 & 65 & 0.05 &  1.030 & 31 & 0.03 &  1.019 & 21 & 0.05 &  1.028 & 22 & 0.02  &  0.986 & 6 & 0.12 \\   
$\rm{D4000}     = (0.84 \pm 0.29) \log \sigma    + \gamma$  &  0.209 & 65 & 0.19 &  0.123 & 31 & 0.11 &  0.166 & 21 & 0.20 &  0.149 & 26 & 0.11  &  0.171 & 7 & 0.20 \\   
\enddata
\tablecomments{(1) Scaling relation. (2) Zero point for the low redshift sample. (3) Number of galaxies
included from the low redshift sample. (4) rms in the Y-direction of the scaling relation for the low redshift sample.
(5), (6), and (7) Zero point, number of galaxies, rms in the Y-direction for the MS0451.6--0305 sample.
(8), (9), and (10) Zero point, number of galaxies, rms in the Y-direction for the RXJ0152.7--1357 sample.
(11), (12), and (13) Zero point, number of galaxies, rms in the Y-direction for the RXJ1226.9+3332 sample.
(14), (15), and (16) Zero point, number of galaxies, rms in the Y-direction for the RXJ0848.6+4453 sample.
Results for Low redshift sample, MS0451.6--0305, RXJ0152.7--1357, and RXJ1226.9+3332 for relations involving Mass, M/L and CN3883 are reproduced from 
J\o rgensen \& Chiboucas (2013). For the $\log {\rm CaHK}$ relation the slope and zero points for the low redshift sample and RXJ0152.7--1357 are from
J\o rgensen et al.\ (2005).}
\tablenotetext{a}{Effective radii from fits with $r^{1/4}$ profiles, $r_{\rm e}=(a_{\rm e}\,b_{\rm e})^{1/2}$.}
\tablenotetext{b}{Effective radii from fits with S\'{e}rsic profiles, $r_{\rm e}=(a_{\rm e}\,b_{\rm e})^{1/2}$.}
\tablenotetext{c}{Effective radii from fits with S\'{e}rsic profiles, $r_{\rm e}=(a_{\rm e}+b_{\rm e})/2$.}
\end{deluxetable*}

\begin{deluxetable*}{llrr}
\tablecaption{Fundamental Plane and Relations for the M/L Ratios \label{tab-FPfit} }
\tablewidth{0pc}
\tabletypesize{\scriptsize}
\tablehead{
\colhead{Cluster} & \colhead{Relation\tablenotemark{a} } & \colhead{$N_{\rm gal}$} & \colhead{rms}
}
\startdata
Coma            & $\log \rm{r_e} = (1.30 \pm 0.08) \log \sigma  - (0.82 \pm  0.03) \log \langle I \rangle _e -0.443 $ & 105 & 0.08 \\  
MS0451.6--0305  & $\log \rm{r_e} = (0.78 \pm 0.18) \log \sigma  - (0.79 \pm  0.11) \log \langle I \rangle _e +0.983 $ &  34 & 0.10 \\  
RXJ0152.7--1357,RXJ1226.9+3332\tablenotemark{b} & $\log \rm{r_e} = (0.65 \pm 0.14) \log \sigma  - (0.67 \pm  0.04) \log \langle I \rangle _e +1.070 $ & 49 & 0.09 \\ 
RXJ0152.7--1357,RXJ1226.9+3332,RXJ0848.6+4453\tablenotemark{c}  & $\log \rm{r_e} = (0.71 \pm 0.20) \log \sigma  - (0.66 \pm  0.06) \log \langle I \rangle _e + \{ 0.911,0.901,0.947\} $ &  57 & 0.09,0.10,0.10 \\  
Coma            & $\log \rm{M/L} = (0.24 \pm 0.03) \log {\rm Mass} -1.754$ & 105 & 0.09 \\  
MS0451.6--0305  & $\log \rm{M/L} = (0.44 \pm 0.09) \log {\rm Mass} -4.499$ &  34 & 0.14 \\  
RXJ0152.7--1357,RXJ1226.9+3332\tablenotemark{b}  & $\log \rm{M/L} = (0.55 \pm 0.08) \log {\rm Mass} -5.845$ &  49 & 0.14 \\  
RXJ0152.7--1357,RXJ1226.9+3332,RXJ0848.6+4453\tablenotemark{c}  & $\log \rm{M/L} = (0.55 \pm 0.06) \log {\rm Mass} -\{ 5.849,5.829,5.911\}$ &  57 & 0.12,0.12,0.15 \\  
Coma            & $\log \rm{M/L} = (1.07 \pm 0.12) \log \sigma -1.560$ & 105 & 0.11 \\  
MS0451.6--0305  & $\log \rm{M/L} = (1.47 \pm 0.29) \log \sigma -2.894$ &  34 & 0.13 \\  
RXJ0152.7--1357,RXJ1226.9+3332\tablenotemark{b}  & $\log \rm{M/L} = (2.26 \pm 0.32) \log \sigma -4.782$ &  49 & 0.17 \\ 
RXJ0152.7--1357,RXJ1226.9+3332,RXJ0848.6+4453\tablenotemark{c}  & $\log \rm{M/L} = (2.16 \pm 0.0.27\log \sigma -\{ 4.542,4.593,4.902\}$ &  57 & 0.12,0.18,0.29 \\  
\enddata
\tablenotetext{a}{The fits for Coma, MS0451.6--0305, RXJ0152.7--1357 and RXJ1226.9+3332 are adopted from J\o rgensen \& Chiboucas (2013) 
and listed here for completeness.}
\tablenotetext{b}{RXJ0152.7--1357 and RXJ1226.9+3332 treated as one sample.}
\tablenotetext{c}{RXJ0152.7--1357, RXJ1226.9+3332, and RXJ0848.6+4453 fit with parallel relations. The zero points and rms for the
three samples are listed in the same order as the clusters.}
\end{deluxetable*}

\subsection{Single stellar population models}

In the analysis we use spectral energy distributions (SED) for 
single stellar population (SSP) models to derive model values of
the absorption line indices that we are able to determine from
the spectra of RXJ0848.6+4453. We use the SEDs from Maraston \& Str\"{o}mb\"{a}ck(2011) 
for a Salpeter (1955) initial mass function (IMF). 
In addition we use mass-to-light (M/L) ratios for very similar models from Maraston (2005).

As in J\o rgensen \& Chiboucas (2013) we have derived model relations linear in
in the logarithm of the age and in the metallicity [M/H]. These relations 
are listed in Table \ref{tab-models} and are used to aid our analysis.
In the fits presented here we include the models with age 1 Gyr as our
sample includes galaxies with very young stellar populations. The fits in
J\o rgensen \& Chiboucas (2013) were derived excluding these very young models. 
Thus, the M/L ratio relation in Table \ref{tab-models} differs slightly
from the relation given by J\o rgensen \& Chiboucas.

The simplest models for the evolution of the stellar populations are passive evolution models. 
In such models it is assumed that after an initial period of star formation
the galaxies evolve passively without any additional star formation. 
The models are usually parameterized by a formation
redshift $z_{\rm form}$, which corresponds to the approximate epoch of the last 
major star formation episode.
In these models the age difference between galaxies at different redshifts is expected
to be equal to the difference in lookback time. Thus, the relations
listed in Table \ref{tab-models} enable us for a given $z_{\rm form}$ to derive the expected offsets
for the M/L ratios and the line indices between the high redshift clusters
and our $z\approx 0$ comparison sample. Alternatively, we can derive a 
formation redshift $z_{\rm form}$ from the measured offsets of the parameters.

The formation redshift may depend on the galaxy properties
and/or the properties of the cluster environment (e.g.\ Thomas et al.\ 2005).
In our analysis we use the model from Thomas et al.
for the high density environment, which prescribes a formation redshift $z_{\rm form}$
dependent on the velocity dispersion of the galaxy.
We convert the velocity dispersion dependency to a mass dependency using
an empirical relation between dynamical mass and velocity dispersion (see J\o rgensen \& Chiboucas 2013).

In our analysis, we implicitly assume that the galaxies we observe
in RXJ0848.6+4453 can be considered progenitors to the galaxies in the clusters 
at lower redshifts.
As discussed in detail by van Dokkum \& Franx (2001) this may not be a valid assumption.
We return to this issue in the discussion (Sect.\ \ref{SEC-DISCUSSION}).

\subsection{Dynamical masses \label{SEC-MASSES} }

The dynamical masses of the galaxies can be determined from the velocity dispersions and effective radii
as ${\rm Mass} = \beta r_e \sigma^2\,G^{-1}$, with $\beta =5$ (Bender et al.\ 1992). 
Cappellari et al.\ (2006) found from integral-field-unit (IFU) data that this approximation provides 
a reasonable mass estimate in the absence of observational data like IFU data, which would enable more 
detailed modeling. 
These authors also tested the use of a coefficient $\beta$ dependent on the S\'{e}rsic index $n_{\rm ser}$.
They derived an expression based on a spherical isotropic model
\begin{equation}
\beta (n_{\rm ser}) = 8.87 - 0.831 n_{\rm ser} +0.0241 n_{\rm ser}^2
\end{equation}
Their conclusion is that this expression does not improve the mass estimate over using $\beta =5$, when
comparing to the masses derived from their full modeling.
Alternatively, van Dokkum et al.\ (2010) derived a fit to the numerical results from Ciotti (1991) 
and suggested the expression
\begin{equation}
\beta (n_{\rm ser}) = 3.316 + 0.026 n_{\rm ser} -0.035 n_{\rm ser}^2 + 0.00172  n_{\rm ser}^3
\end{equation}
We note that this expression does not reach $\beta =5$ for any values of $n_{\rm ser}$ and therefore
is not supported by the detailed modeling of IFU data done by Cappellari et al.\ (2006).
In the following we adopt ${\rm Mass} = 5 r_e \sigma^2\,G^{-1}$ for the mass estimates. 


\subsection{Determination of the star formation rates}

Traditionally determination of the SFR from the [\ion{O}{2}] line has been done from 
the equivalent width, EW[\ion{O}{2}], using the calibrations from Kennicutt (1992) and
Gallagher et al.\ (1989),
\begin{equation}
{\rm SFR} = 1.4 \cdot 10^{-41} L({\rm [O\sc{II}]})
\label{eq-SFR}
\end{equation}
with the SFR in $M_{\sun} yr^{-1}$ and the [\ion{O}{2}] luminosity L([\ion{O}{2}]) in $\rm erg\,sec^{-1}$ 
derived from the equivalent width EW[\ion{O}{2}] as
\begin{equation}
L ({\rm [O\sc{II}]}) = 1.4 \cdot 10^{29} L_B\, {\rm EW}({\rm [O\sc{II}]})
\label{eq-LOII}
\end{equation}
where $L_B$ is the luminosity of the galaxy in rest frame B-band in solar units.

Due to the very faint continuum of most of the star forming galaxies in our sample, the uncertainty on EW[\ion{O}{2}] 
is dominated by the uncertainty on the continuum level and are typically 20-25\%, while the direct measurement 
of the relative [\ion{O}{2}] flux is accurate to 8-10\%. 
Thus, rather than use EW[\ion{O}{2}] and $L_B$ to derive SFR, we instead opt to use the [\ion{O}{2}] flux directly. 
One could argue that because of the limited size spectral aperture
and the possible non-photometric conditions during some of the spectroscopic observations
that we therefore underestimate the total [\ion{O}{2}] flux. To evaluate the size of this effect we derived  $L$[\ion{O}{2}]
both from EW[\ion{O}{2}] and $L_B$ and directly from the  [\ion{O}{2}] flux. 
The [\ion{O}{2}] flux was converted to $L$[\ion{O}{2}] using a luminosity distance of the cluster of 
$D_{\rm L} = 8880.3 {\rm Mpc}$, which corresponds to $z=1.27$ for our adopted cosmology. 
The $L$[\ion{O}{2}] values from the two methods are
compared on Figure \ref{fig-SFRdeterm}. 
The median offset between the two values is $\Delta \log L$[\ion{O}{2}] $=0.1$. Thus, using
the [\ion{O}{2}] flux leads to the SFR being underestimated with a similar amount.
However, this small offset is of no importance for our analysis, and since
the uncertainties 
on $L$[\ion{O}{2}] are about half of those for $L$[\ion{O}{2}] based on the EW[\ion{O}{2}]
we have used Equation \ref{eq-SFR} to derive the SFR.

\begin{figure}
\epsfxsize 8.5cm
\epsfbox{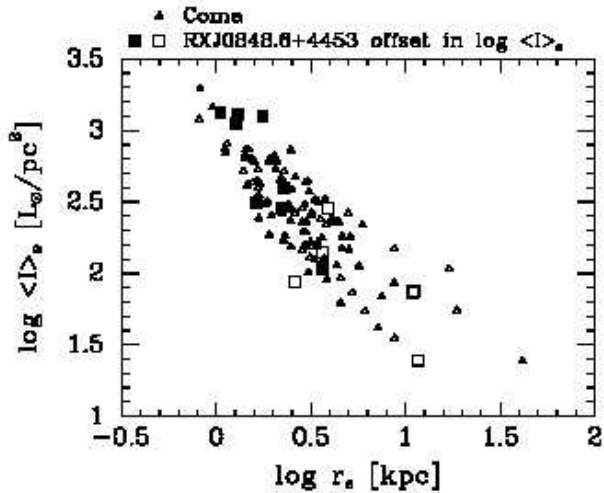}
\caption[]{
Effective radii versus mean surface brightnesses for the Coma cluster sample and the 
RXJ0848.6+4453 sample. 
Open triangles -- Coma cluster;
solid boxes -- bulge-dominated members of RXJ0848.6+4453 with EW[\ion{O}{2}] $\le 5$ {\AA} (sample 5);
open boxes -- bulge-dominated members of RXJ0848.6+4453 with EW[\ion{O}{2}] $> 5$ {\AA} (sample 4).
The data for the RXJ0848.6+4453 sample have been offset with $-0.769$ in 
$\log \langle I \rangle _{\rm e}$ to take into account the luminosity offset relative to the Coma cluster.
After applying this offset, the distribution in the $\log r_{\rm e}$--$\log \langle I \rangle _{\rm e}$ space
of the RXJ0848.6+4453 sample is similar to that of the Coma cluster sample.
\label{fig-lre_lie} }
\end{figure}

\subsection{Sub-samples in RXJ0848.6+4453}

In Table \ref{tab-sample} we divide the 24 members in sub-samples according to available 
spectroscopic parameters, S/N,  S\'{e}rsic index and the strength of the [\ion{O}{2}] emission.
Our main sample consists of sub-samples (4) and (5), which are the bulge-dominated
galaxies with EW[\ion{O}{2}]$> 5${\AA} and $\le 5${\AA}, respectively.

For the analysis of the FP and the relations between masses, sizes and velocity dispersions
we concentrate on sample (5), the bulge-dominated galaxies with  EW[\ion{O}{2}]$\le 5${\AA}. 
Except for allowing galaxies with lower S/N spectra as part of the analysis, 
the selection criteria for sample (5) are the same as used in J\o rgensen \& Chiboucas (2013) as we will
use the results presented in that paper as part of our analysis.
We show the bulge-dominated galaxies with EW[\ion{O}{2}] $> 5${\AA} (sample 4) on the figures, but they are excluded
from the determination of the zero points for the relations.
In the discussion of the absorption line strengths we show samples (4) and (5), as well as the four disk-dominated galaxies from 
sample (2) on the figures. We also show the disk-dominated galaxies from the $z<1$ clusters.
However, the relations as well as zero points are derived from the bulge-dominated galaxies
with EW[\ion{O}{2}]$\le 5${\AA}, only (sample 5 in RXJ0848.6+4453).
In the discussion of the star formation rates derived from
the [\ion{O}{2}] emission all cluster members are included.

Before we proceed, we briefly assess possible selection effects in our RXJ0848.6+4453 sample.
We first note that our sample covers galaxies with effective radii $r_{\rm e}\ge 1$ kpc. This lower
limit is similar to other studies of galaxies at $z\approx 1$, in particular the study by
Saglia et al.\ (2010), with which we will compare in our analysis of the data.
Within the two {\it HST}/ACS fields of RXJ0848.6+4453  there are 33 galaxies with $z_{\rm 850} \le 24.5$
and within 0.1 mag of the CM relation. We have obtained spectroscopy of 20 of these. The 
magnitude distribution of the spectroscopically observed galaxies is not significantly
different from that of all 33 galaxies as tested with Kolmogorov-Smirnov test. 
It is beyond the scope of this paper to determine effective radii and surface brightnesses of
all the galaxies not included in the spectroscopic sample. However, the bulge-dominated members,
(samples 4) and 5) included in the analysis are distributed in the
$\log r_{\rm e}$--$\log \langle I \rangle _{\rm e}$ space similarly to our Coma cluster sample, when the 
luminosity evolution of $-0.769$ in $\log \langle I \rangle _{\rm e}$ is taken into account, see Figure \ref{fig-lre_lie}.
We emphasize that the purpose of this figure is {\it not} to determine the luminosity offset
for RXJ0848.6+4453 relative to the Coma cluster from this projection of the FP, but only to assess
the distributions in $\log r_{\rm e}$ and $\log \langle I \rangle _{\rm e}$.
In conclusion, there are no obvious selection effects related to sizes, luminosities or
surface brightnesses that may bias our results as presented in the following.

\begin{figure*}
\epsfxsize 17cm
\epsfbox{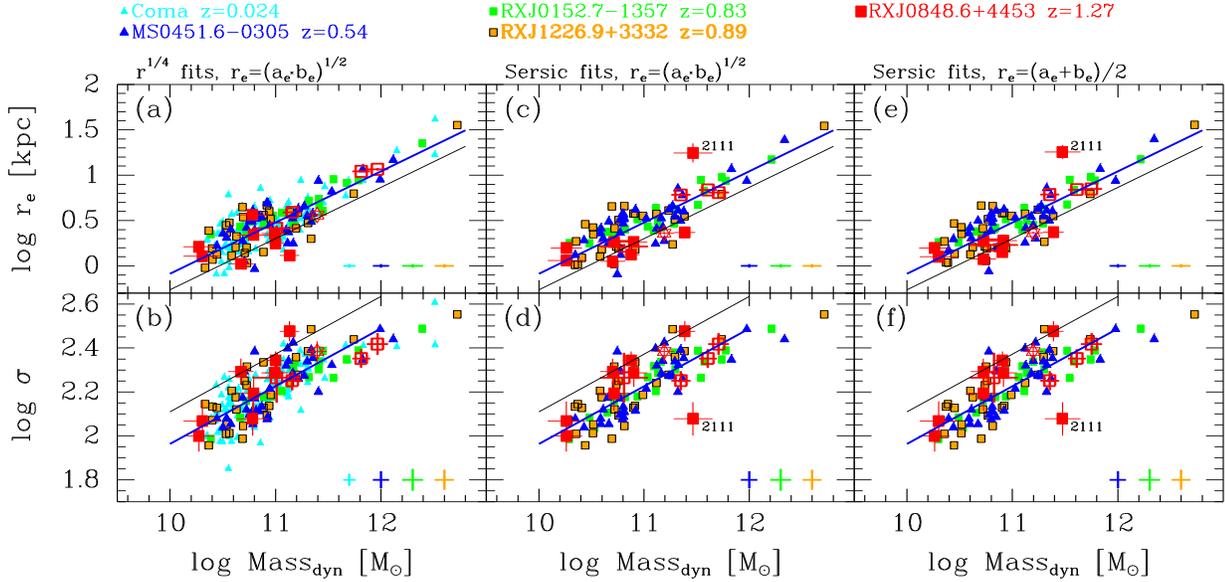}
\caption[]{
Effective radii and velocity dispersions versus dynamical masses. 
Panels (a) and (b) use effective radii from fits with $r^{1/4}$ profiles, $r_{\rm e}=(a_{\rm e}\,b_{\rm e})^{1/2}$.
Panels (c) and (d) use effective radii from fits with S\'{e}rsic profiles, $r_{\rm e}=(a_{\rm e}\,b_{\rm e})^{1/2}$.
Panels (e) and (f) use effective radii from fits with S\'{e}rsic profiles, $r_{\rm e}=(a_{\rm e}+b_{\rm e})/2$.
S\'{e}rsic profile parameters are not available for the Coma cluster galaxies.
Cyan -- Coma cluster members; blue -- MS0451.6--0305; green -- RXJ0152.7--1357; orange -- RXJ1226.9+3332; 
red -- RXJ0848.6+4453.
Open symbols for RXJ0848.6+4453 -- galaxies with EW[\ion{O}{2}]$>$5{\AA}.
Star -- ID 2063, which hosts an AGN.
Blue lines -- best fit relations to the Coma cluster galaxies using  effective radii from fits with $r^{1/4}$ profiles. 
The three clusters at $z<1$ follow the same relation (J\o rgensen \& Chiboucas 2013). 
Black lines -- predicted location of the RXJ0848.6+4453 galaxies under the assumption that the evolution 
found by Saglia et al.\ (2010) is valid for these galaxies.
ID 2111 (marked) has a best fit S\'{e}rsic parameter of $n_{\rm ser}=9.1$ causing the significantly
different position in panels (c)-(f) relative to panels (a) and (b).
\label{fig-lrelmass} }
\end{figure*}

\section{Scaling relations, stellar populations and star formation \label{SEC-RESULTS}}

The main results are described in the following sub-sections. 
Section  \ref{SEC-SIZEMASS} focuses on our 
results regarding the possible size and velocity dispersion
evolution of the galaxies in RXJ0848.6+4453.
Section \ref{SEC-FP} concentrates on the FP for the cluster, while the results regarding
the stellar populations based on measurements of absorption and emission lines are outlined
in Section \ref{SEC-INDICES}.

\begin{figure}
\epsfxsize 8.5cm
\epsfbox{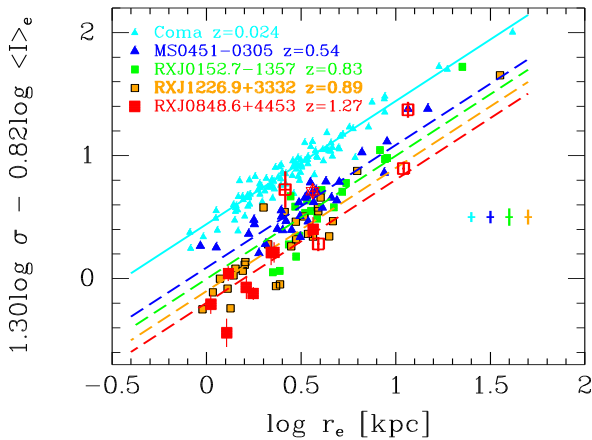}
\caption[]{
The Fundamental Plane shown edge-on.  Symbols as on Fig.\ \ref{fig-lrelmass}.
Cyan line -- best fit relation for the Coma cluster sample.
Dashed lines -- the Coma cluster FP offset to the median zero point for each of the four 
higher redshift clusters. The color coding of the lines match the symbols
(blue -- MS0451--0305; green -- RXJ0152.7--1357; orange -- RXJ1226.9+3332; red -- RXJ0848.6+4453).
\label{fig-FP} }
\end{figure}

\begin{figure*}
\epsfxsize 17cm
\epsfbox{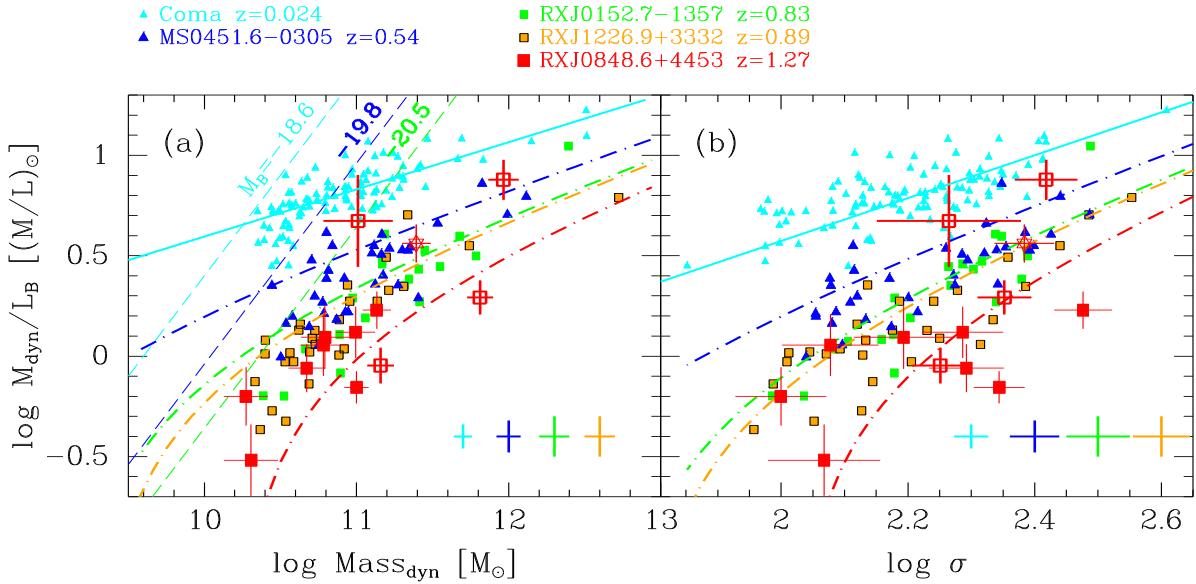}
\caption[]{ 
The dynamical M/L ratios versus the dynamical masses (a) and versus the velocity dispersions (b).
Cyan line -- best fit to the Coma cluster sample. Dot-dashed lines show the predicted location of the 
relations for each cluster redshift based on models for passive evolution
with mass dependent formation redshift (Thomas et al.\ 2005); 
blue --  MS0451--0305; green -- RXJ0152.7--1357; orange -- RXJ1226.9+3332; red -- RXJ0848.6+4453. 
Dashed lines - selection effects.
Symbols as on Fig.\ \ref{fig-lrelmass}.
\label{fig-MLonly} }
\end{figure*}

\begin{figure*}
\epsfxsize 17cm
\epsfbox{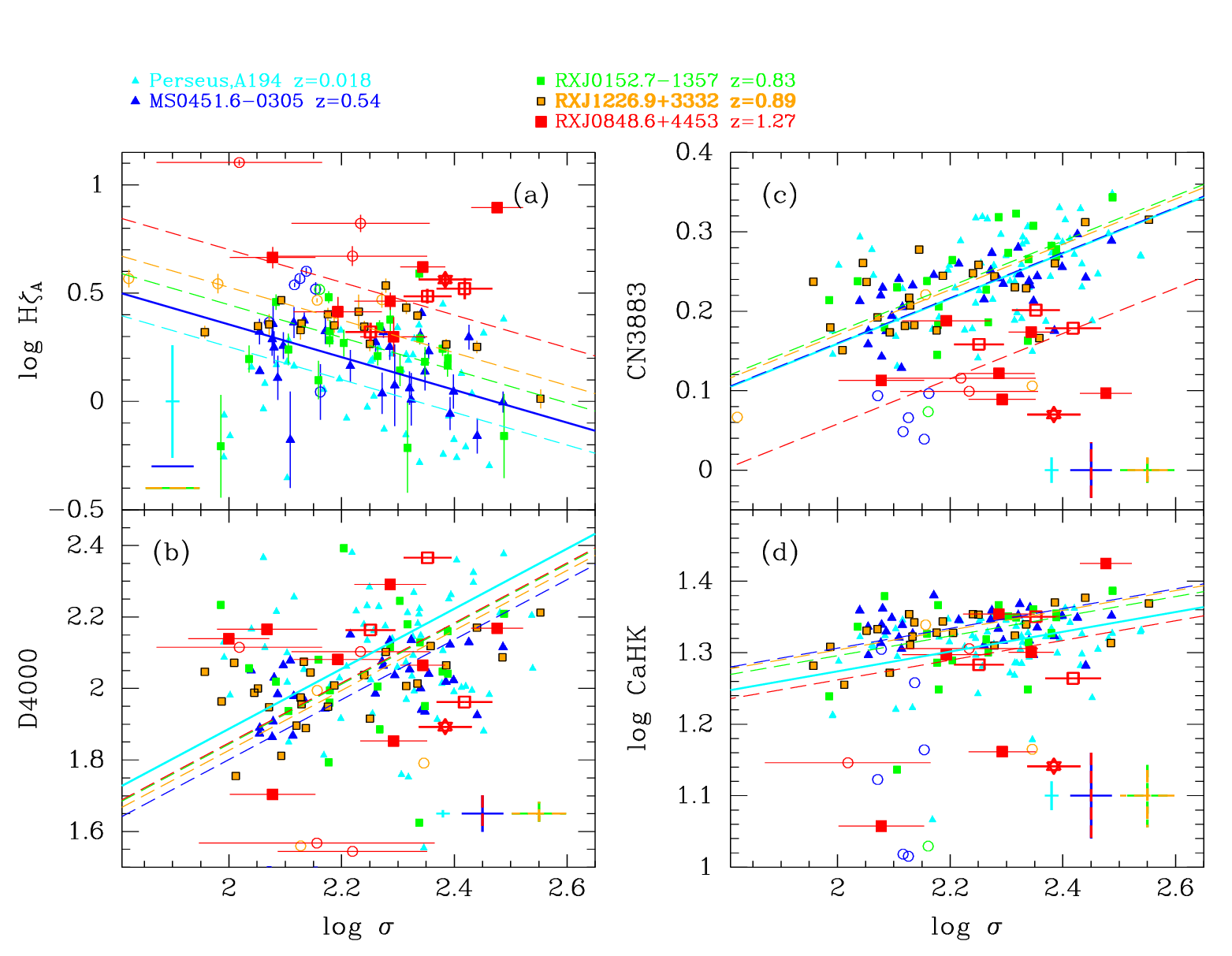}
\caption[]{ 
Absorption line strengths versus velocity dispersions. 
Cyan -- Members of Perseus and A194; blue -- MS0451.6--0305; green -- RXJ0152.7--1357; orange -- RXJ1226.9+3332; 
red -- RXJ0848.6+4453.
Open boxes for RXJ0848.6+4453 -- galaxies with EW[\ion{O}{2}] $>$ 5{\AA}.
Star -- ID 2063, which hosts an AGN.
In addition disk-dominated galaxies ($n_{\rm ser}<1.5$) are shown as open circles, color coded for 
cluster membership as the rest of the symbols.
The solid line on each panel shows the relation for the sample used to establish the slope of the relation, while
the dashed lines show the relations offset to the median zero point for each of the other clusters.
The color coding of the lines match the symbols
(cyan -- low redshift sample; blue -- MS0451--0305; green -- RXJ0152.7--1357; orange -- RXJ1226.9+3332; red -- RXJ0848.6+4453).
For D4000 all samples at $z<1$ were used to establish the slope of the relation, see text.
There are no significant zero point differences with redshift for D4000 and CaHK. For CN3883 only the offset 
of the RXJ0848.6+4453 sample relative to the low redshift sample is significant. All offsets for H$\zeta _{\rm A}$ 
are significant, see text for discussion.
\label{fig-line_sigma} }
\end{figure*}

The established scaling relations are summarized in Tables \ref{tab-relations} and \ref{tab-FPfit},
and shown on Figures \ref{fig-lrelmass}, \ref{fig-FP}, \ref{fig-MLonly}, and \ref{fig-line_sigma}.

\subsection{Radii and velocity dispersions as a function of masses \label{SEC-SIZEMASS} }

Figure \ref{fig-lrelmass} shows the effective radii and velocity dispersions versus the dynamical masses.
The figure shows our data for RXJ0848.6+4453 together with our data for the $z=0.5-0.9$ clusters and 
the Coma cluster sample for effective radii derived from fits with $r^{1/4}$ profiles, panels (a) and (b).
In panels (c)-(d) and (e)-(f) we show RXJ0848.6+4453 together with the $z=0.5-0.9$ clusters 
using S\'{e}rsic effective radii derived as $r_{\rm e} = (a_{\rm e}\,b_{\rm e})^{1/2}$ and 
$r_{\rm e} = (a_{\rm e}+b_{\rm e})/2$, respectively.
On all the panels in Figure \ref{fig-lrelmass} the fit for the Coma cluster sample is shown
(blue lines) together with the prediction of the location of the RXJ0848.6+4453 
sample based on the results from Saglia et al.\ (2010) for dynamical masses (black lines). 
Table \ref{tab-relations} summarizes the zero points for the relations for all the clusters.
For RXJ0848.6+4453 the zero points are derived from the eight bulge-dominated galaxies
with EW[\ion{O}{2}]$\le$5{\AA} (sample 5, see Table \ref{tab-sample}).

As described in J\o rgensen \& Chiboucas (2013), we found no significant evolution in effective
radii or velocity dispersions at a given mass with redshift for the $z=0.5-0.9$ clusters.
In that paper we used effective radii from  $r^{1/4}$ profiles.
As the only cluster in the sample MS0451.6--0305 was observed in a passband matching
roughly rest frame V, rather than rest frame B. Due to color gradients this is expected to result 
in $\log r_{\rm e}$ on average being 0.028 smaller than if measured from a passband matching rest 
frame B (cf., Section \ref{SEC-COMPDATA}). As the dynamical mass also depends on the effective radius, 
correcting for this effect would change the zero point for the size-mass relations for the cluster 
with an insignificant amount of $-0.012$. 

Using the $r^{1/4}$ effective radii, the offsets for the RXJ0848.6+4453 sub-sample of 
bulge-dominated EW[\ion{O}{2}]$\le$5{\AA} galaxies relative to the Coma cluster relations is 
about a third of what is expected based on the results from Saglia et al. 
Further, the offsets are significant only at the 1-sigma level.
If we use the effective radii from the S\'{e}rsic profiles the offsets relative to the $z=0.5-0.9$
clusters are slightly larger, though still only significant at the 1-1.5 sigma level.
We note that Saglia et al.\ used effective radii derived as $r_{\rm e} = (a_{\rm e}+b_{\rm e})/2$.
However, we find no significant difference between using $r_{\rm e} = (a_{\rm e}\,b_{\rm e})^{1/2}$
and $r_{\rm e} = (a_{\rm e}+b_{\rm e})/2$ for our samples.

Concentrating on the results based on the  $r^{1/4}$ effective radii, we convert the
the median offsets in radii and velocity dispersions to an estimate of the median change of
the dynamical mass, using  ${\rm Mass} = 5 r_e \sigma^2\,G^{-1}$. This indicates in an insignificant 
mass change, $\Delta \log {\rm Mass} = 0.01 \pm 0.11$. 
Taking the uncertainty into account we can interpret this as an upper limit on the mass increase 
of $\approx 23$ percent.

The five bulge-dominated emission line galaxies (sample 4 in Table \ref{tab-sample}) show no offsets in 
effective radii or velocity dispersion relative to the Coma cluster relations (Fig.\ \ref{fig-lrelmass})
when using  $r^{1/4}$ effective radii and also no offsets relative to the $z=0.5-0.9$ clusters
when using the effective radii based on S\'{e}rsic profiles. 

\subsection{The Fundamental Plane and relations for the M/L ratios \label{SEC-FP} }

Figure \ref{fig-FP} shows the FP edge-on for the RXJ0848.6+4453 samples (4) and (5)
(Table \ref{tab-sample}) together with our samples for $z=0.5-0.9$ clusters and the Coma cluster samples. 
On Figure \ref{fig-MLonly} we show the FP as the M/L ratios versus the dynamical masses and versus
the velocity dispersions.
Tables \ref{tab-relations} and \ref{tab-FPfit} summarize the derived relations and zero points. 
All results in these two tables relating to only the $z=0.5-0.9$ clusters and the Coma cluster samples
are adopted from J\o rgensen \& Chiboucas (2013) and reproduced here to aid the discussion of the
results for the RXJ0848.6+4453 sample.

The RXJ0848.6+4453 sample appears to follow a steep relation in M/L versus Mass and M/L versus velocity dispersion, 
as is the case for the two highest redshift clusters RXJ0152.7--1357 and RXJ1226.9+3332 from J\o rgensen \& Chiboucas.
However, because the sample contains only eight galaxies it is not possible to determine the slopes
of the relations or the coefficients in the FP by fitting only this sample. Instead
we have determined the best fit relations by fitting parallel relations to the three clusters.
Thus, the assumption is that only the zero point varies between these clusters. 
Table \ref{tab-FPfit} summarizes the derived relations.
As the zero points for RXJ0152.7--1357 and RXJ1226.9+3332 are not significantly different from each
other (cf.\ J\o rgensen \& Chiboucas 2013) we use their common zero point.
We then derive the zero point difference for RXJ0848.6+4453 relative to that, and from there
derive the formation redshift $z_{\rm form}$ required for the galaxies in RXJ0848.6+4453 to
evolve passively to the location of the M/L versus Mass relation for RXJ0152.7--1357 and RXJ1226.9+3332.
However, the zero point difference is so small that $z_{\rm form} = \infty$ is needed.
Thus, we conclude that the location of the M/L versus Mass relation for RXJ0152.7--1357 and RXJ1226.9+3332
cannot be the result of passive evolution of a higher redshift sample like the RXJ0848.6+4453 sample.

The zero point offset for RXJ0848.6+4453 (sample 5) relative to the Coma cluster sample
is consistent with passive evolution and a formation redshift of
$z_{\rm form}= 1.95^{+0.22}_{-0.15}$. All eight galaxies in this sample have $\log {\rm Mass}<11.1$.
The higher mass galaxies in the cluster have stronger emission lines and show a much larger scatter around
the M/L versus Mass relation. 
We note that the $z=0.5-0.9$ clusters contain low mass galaxies with $z_{\rm form}$ significantly below 1.95.
We return to the discussion of this in Section \ref{SEC-DISCUSSION}.

\subsection{Stellar populations: Absorption and emission lines \label{SEC-INDICES} }

In Figure \ref{fig-line_sigma} we show the available absorption line indices versus the velocity dispersions. 
The figure includes the disk-dominated galaxies from all the clusters. For RXJ0848.6+4453 these are
the galaxies listed in Table \ref{tab-sample} as sample (2).
Table \ref{tab-relations} lists the relations shown on the figure.
The correlation between the H$\zeta _{\rm A}$ indices and the velocity dispersions is only significant
for the sample of galaxies in MS0451.6--0305, for which a Kendall's $\tau$ rank order test gives a probability
of 0.7\% of no correlation being present. For all the other cluster samples, the probabilities are 22\% or 
larger that there is no correlation. We have therefore established the slope of the relation fitting only
the MS0451.6--0305 sample. For this relation the residuals were minimized in the direction of H$\zeta _{\rm A}$.
The zero points for all the samples were then derived using the slope established from the MS0451.6--0305 sample.

The correlation between the D4000 indices and the velocity dispersions for the individual clusters is weak. However,
a Kendall's $\tau$ rank order test for the joint sample of all the galaxies in the $z<1$ samples
yields a probability of 0.4\% of no correlation being present. Thus, we determine the slope by fitting 
the full sample. After this, we determine the cluster specific zero points relative to this relation.

The main results we derive from Figure \ref{fig-line_sigma} and the relations in Table \ref{tab-relations} are 
(1) the presence of very strong H$\zeta$ lines in 
the RXJ0848.6+4453 galaxies compared to the lower redshift samples, (2) the RXJ0848.6+4453 galaxies
have significantly weaker CN3883 than found for the lower redshift samples, and (3)
the RXJ0848.6+4453 galaxies are not significant offset in D4000 or CaHK relative to the lower redshift samples, though
the scatter in these indices may be somewhat larger than for the lower redshift samples.
It should be noted that except for CN3883 being weaker in the RXJ0848.6+4453 galaxies than in the 
lower redshift samples, none of the offsets in D4000, CaHK and CN3883 between the $z=0.5-1.3$ clusters
and the low redshift comparison sample are significant. We return to this in the discussion in
Section \ref{SEC-DISCUSSION}.

\begin{figure*}
\epsfxsize 17cm
\epsfbox{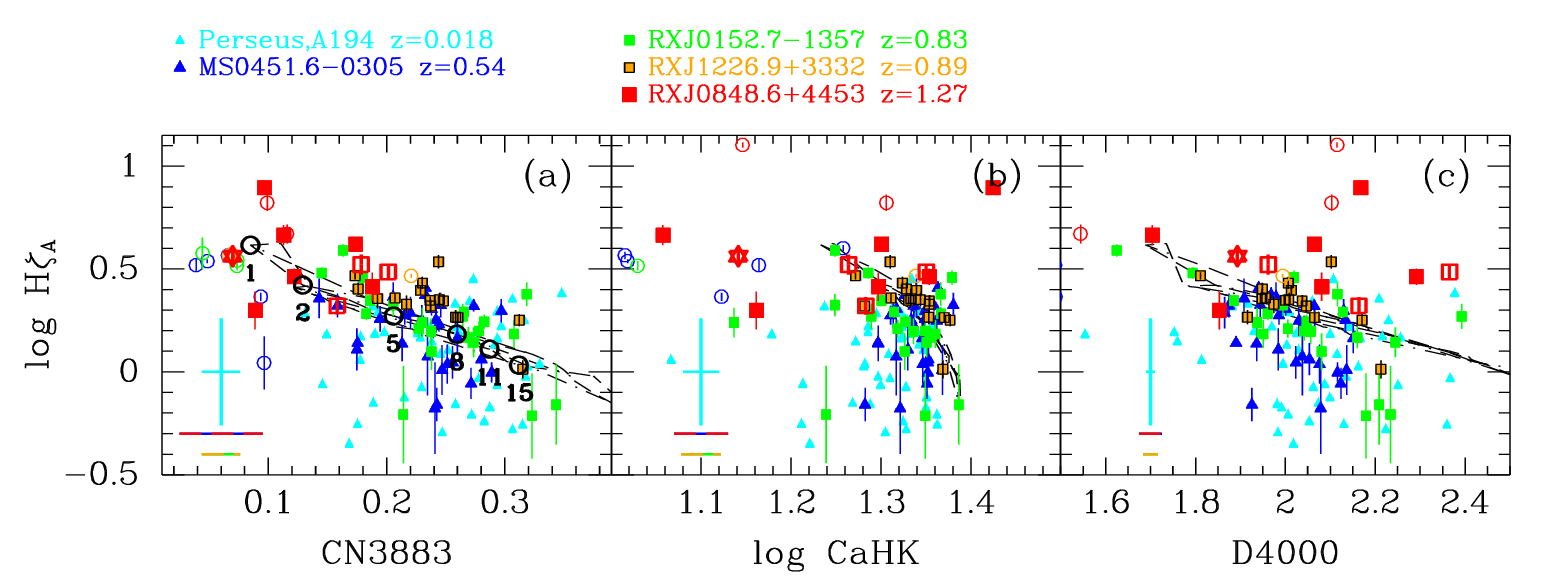}
\caption[]{ 
Absorption line strengths versus each other. 
Symbols as on Fig.\ \ref{fig-line_sigma}.
Black dashed lines -- model values based on from SSPs from Maraston \& Str\"{o}mb\"{a}ck (2011).
The models are degenerate in age and metallicity [M/H]. On panel (a) the black open circles correspond
to models with solar [M/H] and ages of 1, 2, 5, 8, 11, and 15 Gyr as labeled below the points, see text for details.
\label{fig-line_line} }
\end{figure*}

We then investigate whether it is possible to use a combination of indices to estimate
ages and metallicities of the galaxies in RXJ0848.6+4453.
Figure \ref{fig-line_line} shows the strength of H$\zeta _{\rm A}$ versus CN3883, CaHK, and D4000.
The figure also shows model predictions based on the SSP SEDs from Maraston \& Str\"{o}mb\"{a}ck (2011). 
These predictions are degenerate in metallicity and age. 
Thus, we can only approximately estimate the ages of the stellar populations from the indices. 
The metallicities cannot be constrained beyond noting that due to the strength of CaHK and CN3883 the metallicities
of the RXJ0848.6+4453 galaxies must be similar to those of the lower redshift galaxies, ie.\ solar or above solar.
The best age estimates come from using H$\zeta _{\rm A}$ versus CN3883.
The area in this diagram populated by the RXJ0848.6+4453 galaxies (see Fig.\ \ref{fig-line_line}a) can only be reached if the
stellar population ages are 1--2 Gyr.
Only a handful of the $z=0.5-0.9$ bulge-dominated galaxies populate the same part of the diagram.

Figure \ref{fig-Rcl_SFR} shows the star formation rates derived from the [\ion{O}{2}] emission
as well as the strength of the H$\zeta$ line versus the cluster center distance, $R_{\rm cluster}$, 
and versus the dynamical mass.
As not all galaxies have determination of the dynamical mass, Figure \ref{fig-Rcl_SFR}b contains
fewer points than  Figure \ref{fig-Rcl_SFR}a.
Figures \ref{fig-Rcl_SFR}c and d show only galaxies with measurements of H$\zeta _{\rm A}$, which 
in particular excludes the three low mass (Mass $< 10^{10.5} {\rm M}_{\sun}$) galaxies 
that are included on Figure \ref{fig-Rcl_SFR}b.
From Figures \ref{fig-Rcl_SFR}a and c we conclude that star forming galaxies and galaxies with
strong H$\zeta _{\rm A}$ are present throughout the 
cluster. There is no trend in SFR or H$\zeta _{\rm A}$ with $R_{\rm cluster}$, and also no radius within which
star formation is absent.
Figures \ref{fig-Rcl_SFR}b and d show that star formation and strong H$\zeta _{\rm A}$ are 
present in the most massive of the bulge-dominated galaxies. 
However, the SFR is well below that of galaxies on the star formation ``main sequence'' (e.g., Wuyts et al.\ 2011).

\section{Discussion \label{SEC-DISCUSSION} }

In this discussion, we compare the results regarding the size and velocity dispersion evolution to
previous results and discuss the possible effect of the cluster environment on this evolution.
We then discuss the evolution in the stellar population as reflected through
the M/L versus mass relation, the absorption line strengths and the [\ion{O}{2}] emission.

\subsection{Size and velocity dispersion evolution \label{SEC-DISCSIZE}}

In our study of the massive $z=0.5-0.9$ clusters also included in this paper we
found no indication of evolution with redshift in sizes or velocity dispersions of the passive bulge-dominated 
galaxies (J\o rgensen \& Chiboucas 2013).
Our results on RXJ0848.6+4453 presented here indicate a very small (1-sigma) difference between the 
sizes and velocity dispersions of galaxies at a given mass in this cluster compared to those at $z\approx 0$.
A similar conclusion regarding the size evolution was reached for 16 galaxies in RXJ0848.6+4453
by Saracco et al.\ (2014), who using stellar masses put an upper limit on the 
size evolution of $r_{\rm e} \propto (1+z)^{-0.1}$ equivalent to $\approx 8$\% from $z=1.27$ to the present.

Recent studies have addressed the question of the possible environmental dependence
of the size and velocity dispersion evolution of galaxies. 
Several authors have directly compared sizes of passive galaxies in high density and low density environments.
Lani et al.\ (2013) finds that at $z>1$ galaxies in high density environments are $\approx 50$ percent larger
than those in low density environments. Delaye et al.\ (2013) reach a similar conclusion. 
Converting the Delaye et al.\ result on size dependences on redshift in the field and in clusters
to a difference in sizes at $z \approx 1.3$ gives a size difference of $\approx 30$ percent, with the 
cluster galaxies being larger.
Specifically for clusters Delaye et al.\ find $r_{\rm e} \propto (1+z)^{-0.53}$ using stellar masses, 
which is in agreement with the result from Saglia et al.\ (2010) using dynamical masses for 
the clusters in the EDisCS survey. 
At $z \approx 0$ any differences in sizes between field and cluster galaxies appear to have 
disappeared, or at least become undetectable, see Huertas-Company et al.\ (2013).

In a study of a $z\approx 1.8$ cluster, Newman et al.\ (2014) on the other hand find no difference between
the sizes of the galaxies in that cluster and field galaxies at similar redshifts. Thus, contradicting
the above results and arguing that these other results are due to differences in the morphological 
mixture of the samples studied.

\begin{figure}
\begin{center}
\epsfxsize 6.8cm
\epsfbox{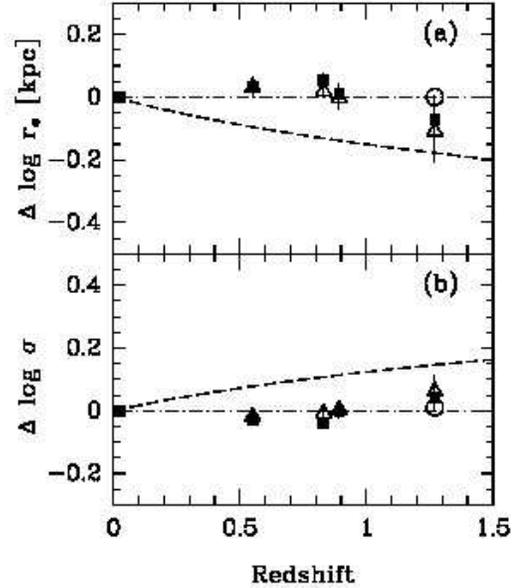}
\end{center}
\caption[]{
Zero point offsets for size-Mass and velocity dispersion-Mass relations as a function of cluster redshifts.
Solid boxes -- based on effective radii from $r^{1/4}$ profile fits; triangles -- based on effective radii
from fits with S\'{e}rsic profiles; open circles for RXJ0848.6+4453 only -- offset relative to the Coma cluster 
sample when galaxies younger than $\approx 8.3 $Gyr are removed from the Coma cluster sample.
Dashed lines -- results from Saglia et al.\ (2010);
dot-dashed lines -- reference for no evolution.
Without correcting for progenitor bias due to the presence of young galaxies in the Coma cluster sample, 
RXJ0848.6+4453 at $z=1.27$ shows roughly a third the evolution with redshift compared to the 
results from Saglia et al.  
With correction for progenitor bias, there is no evolution with redshift, see text for discussion.
\label{fig-lrelmasszp} }
\end{figure}

Of these results, only Saglia et al.\ (2010) use dynamical masses and their result is therefore most directly comparable to
our result and also the only study that make it possible to directly compare to our results for the
evolution of the velocity dispersion. 
On Figure \ref{fig-lrelmasszp} we summarize the results as the median offset for each of the $z=0.5-1.3$ clusters relative to the location
of the Coma cluster relations. The dashed lines on this figure shows the results from 
Saglia et al.\ for dynamical masses.  The size and velocity dispersion evolution required
to bring our RXJ0848.6+4453 to the $z\approx 0$ location of the relations with mass is only about
a third of that found by Saglia et al. Formally the zero point differences relative to the Coma 
cluster sample are only significant at the one-sigma level.

The result from Saglia at al.\ is corrected for progenitor bias using a method adopted from
Valentinuzzi et al.\ (2010).
We have chosen to assess the effect of progenitor bias on our result in an empirical fashion using 
the information about the H$\beta$ line strength for our Coma cluster sample to evaluate the ages
of galaxies in this sample. Using the model relation between the H$\beta$ line strength,
age, metallicity [M/H] and abundance ratio [$\alpha$/Fe] based on stellar population models
from Thomas et al.\ (2005) and established in J\o rgensen \& Chiboucas (2013) we can remove galaxies
from the Coma cluster sample too young to have progenitors in our RXJ0848.6+4453 sample.
The difference in lookback time between the two clusters is 8.3 Gyr for our adopted cosmology.
With [M/H]=0.3 and [$\alpha$/Fe]=0.3 as typical average values for galaxies in the Coma cluster 
sample (see J\o rgensen \& Chiboucas 2013), we then require $\log {\rm H}\beta _{\rm G}\le0.29$
in order for the galaxies to be older than about 8.3 Gyr.
Doing so reduces the zero point difference between the Coma cluster sample and our sample in
RXJ0848.6+4453 for the size-mass relation to zero, while the offset for the velocity dispersion--mass
relation is 0.011 in log space. These values are shown on Figure \ref{fig-lrelmasszp} as open circles.
Thus, the correction for progenitor bias decreases the possible evolution in size and 
velocity dispersions, as also found by Saglia et al.

Our result for  RXJ0848.6+4453 combined with our previous results for the massive $z=0.5-0.9$ 
clusters (J\o rgensen \& Chiboucas 2013; also shown on Fig.\ \ref{fig-lrelmasszp}) indicates 
an even larger difference between the evolution in dense environments and in the field
than found by, e.g., Delaye et al.\ (2013). 
However, we note that the clusters in our $z<1$ sample are 
significantly more massive than those in the EDisCS sample. 
Based on models for cluster mass evolution (van den Bosch 2002, see Fig.\ \ref{fig-M500}) 
RXJ0848.6+4453 is expected to evolve to a similarly massive cluster by the present time.
Of the above studies, we can only confirm that Delaye et al.\ include some similarly massive 
clusters. 
Further, our galaxy samples in the clusters are selected consistently based on both morphology
$n_{\rm ser}>1.5$ and spectroscopic properties (passive galaxies with EWOII$\le$5 {\AA}). 
Therefore, our result cannot be explained as due to a mix-up of sample selections.

We conclude that with homogeneous selection of samples in very massive clusters, and use
of dynamical galaxy masses, we find that a size and velocity dispersion evolution from $z \approx 1.3$ 
to the present of $\approx 16$\% and $\approx 7$\%, respectively. Both differences are 
significant only at the one-sigma level.
Further, the evolution is likely to have completed by $z\approx 0.9$.
We put an upper limit of 23\% on the increase in mass associated with this evolution (see Sect.\ \ref{SEC-SIZEMASS}).

\begin{figure}
\begin{center}
\epsfxsize 8.5cm
\epsfbox{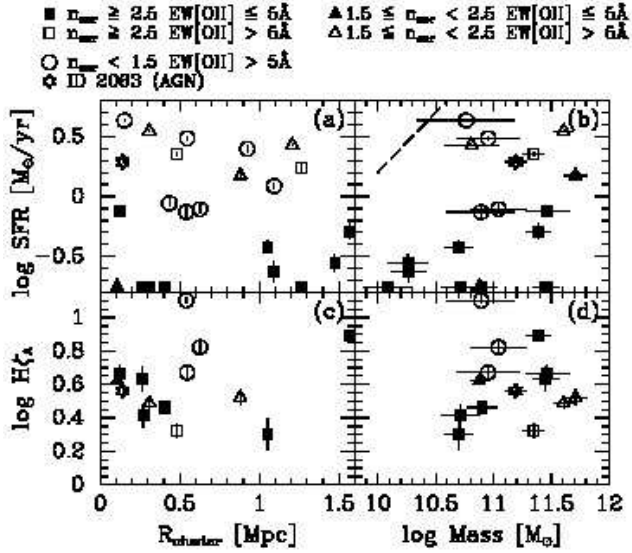}
\end{center}
\caption[]{ 
Star formation rates (SFR) and $\log {\rm H}\zeta _{\rm A}$ versus cluster center distances $R_{\rm cluster}$
(panels a and c) and versys the dynamical masses of the galaxies (panels b and d)
The figures shows data for confirmed members of RXJ0848.6+4453.
Solid symbols -- EW[\ion{O}{2}] $\le 5${\AA}; open symbols -- EW[\ion{O}{2}] $> 5${\AA}.
Squares -- $n_{\rm ser}\ge 2.5$; triangles $1.5 \le n_{\rm ser} < 2.5$; circles -- $n_{\rm ser} < 1.5$; 
star -- ID 2063, which hosts an AGN, see text.
Dashed line on panel (b) - ``main sequence'' of star formation at $z=0$ (Wuyts et al.\ 2011).
\label{fig-Rcl_SFR} }
\end{figure}

\begin{figure*}
\begin{center}
\epsfxsize 15cm
\epsfbox{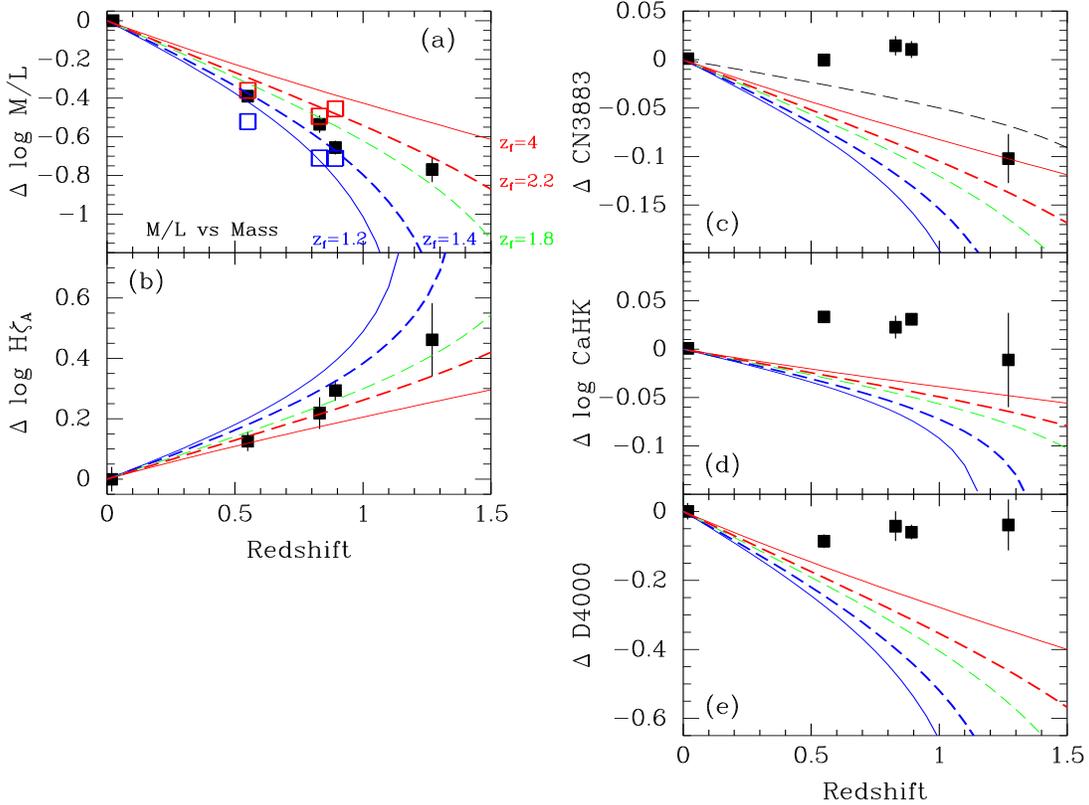}
\end{center}
\caption[]{
The zero point offsets of the scaling relations for the $z=0.5-1.3$ cluster samples  relative to the low redshift samples, shown as 
a function of redshift. Results for the $z<1$ clusters for the M/L ratios and CN3883 are adopted from J\o rgensen \& Chiboucas (2013).
Predictions from models for passive evolution based on models 
from Maraston (2005) and Maraston \& Str\"{o}mb\"{a}ck (2011)
are overplotted, labeled with the assumed formation redshift $z_{\rm form}$.
Black dashed line on panel (c) -- the prediction for $z_{\rm form}=1.8$ if adopting the 
age dependence of CN3883 discussed in J\o rgensen \& Chiboucas (2013).
Black points -- median for the full sample in each cluster; 
red points -- for $z-0.5-0.9$ clusters, galaxies with $\log {\rm Mass} \ge 11$; 
blue points -- for $z-0.5-0.9$ clusters, galaxies with $\log {\rm Mass} < 11$. 
\label{fig-zp} }
\end{figure*}

\subsection{Stellar population evolution \label{SEC-DISCPOPS}}

Figure \ref{fig-zp} summarizes the changes in the M/L ratios and the absorption
line strengths as the zero point offsets of the $z=0.5-1.3$ cluster samples relative to the low redshift samples.
All clusters are shown in these figures. Results for the $z=0.5-0.9$ clusters 
for M/L and CN3883 are adopted from J\o rgensen \& Chiboucas (2013).

We compare the zero point for the M/L-Mass relation for the RXJ0848.6+4453 bulge-dominated
galaxies with EW[\ion{O}{2}]$\le 5${\AA} to that of the Coma cluster sample. 
The difference is consistent with passive evolution and a formation redshift
of $z_{\rm form}=1.95^{+0.22}_{-0.15}$. This is higher than for the
$z=0.5-0.9$ clusters (J\o rgensen \& Chiboucas 2013, see also Fig.\ \ref{fig-zp}a)
for which we found $z_{\rm form}\approx 1.4$.
The difference becomes even larger, when considering that the
eight passive galaxies in our RXJ0848.6+4453 all have masses below $10^{11.1} {\rm M}_\sun$.
For similar low-mass galaxies in the $z=0.5-0.9$ clusters the formation redshift
is $\approx 1.2$ (blue points on Fig.\ \ref{fig-zp}a).

We speculate that the reason for this difference is that a large fraction
of the low mass galaxies in $z=0.5-0.9$ clusters have entered the passive
bulge-dominated population more recently than $z\approx 1.3$. Thus, the difference is due to 
a progenitor bias. The RXJ0848.6+4453 sample does not include the progenitors of those young
bulge-dominated galaxies in the $z=0.5-0.9$ clusters, as such progenitors
observed at $z\approx 1.3$ would have very strong star formation and maybe also
not yet be bulge-dominated. 
This conclusion is similar to that reached by S\'{a}nchez-Bl\'{a}zquez et al.\ (2009)
who argue that 50\% of the low mass galaxies on the red sequence 
at low redshift have entered the red sequence recently.

Turning to the strength of the H$\zeta$ line, our results support the 
idea that RXJ0848.6+4453 experienced a major episode of star formation 1-2 Gyr prior,
consistent with $z_{\rm form}=1.95$. The strengths of  H$\zeta$ seen in the 
RXJ0848.6+4453 galaxies can according to stellar population models only be achieved
for such young stellar populations (see Fig.\ \ref{fig-line_line}a).
Within the uncertainty, this is also consistent with the zero point offset of the H$\zeta _{\rm A}$--velocity dispersion
relation for RXJ0848.6+4453 relative to the low redshift sample (Fig.\ \ref{fig-zp}b).
The occurrence of such a star formation episode is further supported by the fact
that the most massive bulge-dominated galaxies in the cluster all have significant
[\ion{O}{2}] emission.

It is worth noting that all the bulge-dominated galaxies show young stellar populations
either through the strength of the H$\zeta$ line or presence of [\ion{O}{2}] emission, or both.
These galaxies are distributed throughout the cluster (see Fig.\ \ref{fig-Rcl_SFR}).
Thus, there is no indication that the star formation has yet been quenched in the center
of the cluster. This is different from results for the more massive clusters 
RDCS\,J1252.9--2927 ($z=1.24$) and XMMU\,J2235.3--2557 ($z=1.4$) at similar redshifts.
Both of these clusters show an absence of star forming galaxies in the very center
of the clusters (Nantais et al.\ 2013; Gr\"{u}tzbauch et al.\ 2012).
We speculate that the difference is related to the mass of the clusters, as 
these two latter clusters have 
$M_{500}=6.1 \times 10^{14} {\rm M}_\sun$ (Stott et al.\ 2010) and
$M_{500}=4.4 \times 10^{14} {\rm M}_\sun$ (Rosati et al.\ 2009, see also Stott et al.\ 2010), respectively. 
Thus, masses that are a factor 3--4.5 larger than the mass of RXJ0848.6+4453.
Lynx E, which is located in the same supercluster as RXJ0848.6/4453/Lynx W, and is similarly
massive as RDCS\,J1252.9--2927 and XMMU\,J2235.3--2557 ($M_{500}=4.7 \times 10^{14} {\rm M}_\sun$, Stott et al.\ 2010)
has not yet been studied in the same detail.

Two studies of $z\approx 2$ clusters find similar results regarding the presence of star
formation in massive cluster galaxies as we find for  RXJ0848.6+4453. 
Strazzullo et al.\ (2013) find in their study of J1449+0856 ($z=2.0$) that the cluster hosts
massive star forming as well as passive galaxies in the core. This cluster has an estimated mass of
$M_{\rm 500} = 3.4 \times 10^{13}{\rm M}_\sun$  
(Gobat et al.\ 2011, with the conversion $M_{\rm 200} \approx 1.54 M_{\rm 500}$ from Brodwin et al.\ 2011).
Thus, the cluster is significantly less massive than RXJ0848.6+4453, but based on the models
for cluster mass evolution with redshift (see Fig.\ \ref{fig-M500}) it is expected to evolve into a cluster mass similar to that of
RXJ0848.6+4453 at $z\approx 1.3$. 

Tanaka et al.\ (2013) investigated the cluster galaxies around the radio source PKS1138--262 ($z=2.16$) 
and conclude that this cluster also contains a mix of massive star forming and passive galaxies, 
as if the cluster is in the process of quenching the star formation.
Shimakawa et al.\ (2014) derived the velocity dispersion of H$\alpha$ and Ly$\alpha$ emitters
that they consider part of the virialized core of the cluster. 
They find $\sigma _{\rm cluster}=683 {\rm km\,s^{-1}}$ from which they derive 
$M_{\rm 200} = 1.71 \times 10^{14}{\rm M}_\sun$, or $M_{\rm 500} = 1.1 \times 10^{14}{\rm M}_\sun$
with the conversion from Brodwin et al.
This cluster mass represents an upper limit as even the core may not yet be virialized.
However, it appears that the cluster may already have a mass comparable to that of  RXJ0848.6+4453
and therefore will be expected to evolve into a significantly more massive cluster than
RXJ0848.6+4453 at later epochs.

The difference in lookback time between $z\approx 2$ and $z\approx 1.3$ is about 1.5 Gyr with our 
adopted cosmology. Thus, it is plausible that we are in fact seeing  J1449+0856, PKS1138--262 and
RXJ0848.6+4553 at slightly different epochs during the quenching of the star formation as the galaxies
fall into the dense cluster environment. We speculate that this process takes place earlier
(or faster) in more massive clusters like RDCS\,J1252.9--2927 and XMMU\,J2235.3--2557.
Detailed spectroscopic observations of both those two clusters and higher redshift progenitors,
of such massive clusters are needed to resolve this question. Based on its mass estimate
PKS1138--262 may be such a progenitor.

The results for the M/L ratios, H$\zeta$ line strength and [\ion{O}{2}] emission appear to
give a consistent picture of the stellar populations in the bulge-dominated RXJ0848.6+4453
galaxies and an epoch of the period of the last star formation of $z_{\rm form}=1.95$. 
However, the strengths of the metal lines CN3883 and CaHK, as well as the D4000 strength
appear in contradiction with these results. These three indices are all stronger than expected for 
1-2 Gyr old stellar populations with metallicities of [M/H] $= 0-0.3$, based on the predictions from SEDs from
Maraston \& Str\"{o}mb\"{a}ck (2011).
This is shown in Figure \ref{fig-zp}c-e, where we show the zero point offsets
for these three indices relative to our low redshift sample. The figure also shows
the results for the $z=0.5-0.9$ clusters from J\o rgensen \& Chiboucas (2013).
As our RXJ0848.6+4453 sample is the first sample of $z>1$ galaxies with measured metal absorption 
line indices, we cannot compare this result to any prior results. 
We do caution that the stellar population models may not 
correctly model these blue indices. In J\o rgensen \& Chiboucas (2013) we showed that
the SEDs from Maraston \& Str\"{o}mb\"{a}ck predict CN3883 too strong for a given CN$_2$ index.
Because of that, the prediction of the age dependency for CN3883 based on these
SEDs is also significantly stronger than if we adopt the 
dependency derived from the CN$_2$ index as we did in J\o rgensen \& Chiboucas (2013).
On Figure \ref{fig-zp}c we show this latter prediction as well for $z_{\rm form}=1.8$.
From this is it is clear that it is yet not straight forward to interpret the 
strength of this blue metal index within the frame work of the SSP models.
Thus, to make progress on the interpretation of the metal indices, further progress
is needed on the modeling. This is beyond the scope of this paper.
Additional data for $z>1$ cluster galaxies are also needed to confirm
the presence of these strong metal lines in galaxies at these redshifts.

\section{Conclusions \label{SEC-CONCLUSION}}

We have used deep ground-based optical spectroscopy from Gemini North and {\it HST}/ACS imaging to investigate the
structure and stellar populations of galaxies in RXJ0848.6+4453/Lynx W at redshift $z=1.27$.
Our main conclusions are as follows:

\begin{enumerate}
\item
At a given dynamical mass, the galaxies in RXJ0848.6+4453 show only
a very small difference in size and velocity dispersion when compared to
our low redshift sample. Formally the effects are at the one-sigma level
and about one third of that expected if extrapolating
the results from the EDisCS survey (Saglia et al.\ 2010).
Our result adds support to the idea that the evolution of sizes and 
velocity dispersions depends on the cluster environment and is 
accelerated in high mass clusters compared to poorer clusters and the field.

\item
The bulge-dominated galaxies in  RXJ0848.6+4453 populate a Fundamental
Plane (FP) similar to that seen for lower redshift galaxies. The slope for 
the  RXJ0848.6+4453 is similar to that found for $z=0.8-0.9$ clusters, though
the sample is too small for an independent determination of the slope.
The FP zero point is in agreement with a model of passive evolution with
a formation redshift of $z_{\rm form}=1.95^{+0.22}_{-0.15}$.
This is a higher $z_{\rm form}$ than we previously found for our sample of
galaxies in $z=0.8-0.9$ clusters at similar galaxy masses (J\o rgensen \& Chiboucas 2013).
Our result show that the low mass end of the FP is populated already
at $z\approx 1.3$, but also that additional passive galaxies are added at 
later epochs.

\item
The bulge-dominated galaxies in RXJ0848.6+4453 have very strong H$\zeta$ absorption
lines and the highest mass bulge-dominated galaxies also contain significant
[\ion{O}{2}] emission. From both of these facts, we conclude that the galaxies have 
experienced an episode of star formation about 1-2 Gyr prior to the epoch
equivalent to the cluster redshift. This is in agreement with the
formation redshift determined from the FP.
The data indicate that this episode of star formation was widespread
in the cluster, and that star formation has not yet been fully quenched in the
very center of the cluster.

\item
The metal lines CN3883 and CaHK, as well as D4000 are stronger than
expected if these galaxies have $z_{\rm form}=1.95$ and are to passively
evolve into galaxies similar to those in our lower redshift samples.
Further investigation of metal lines in $z>1$ galaxies are needed to
shed light on this apparent contradiction with the results from the 
FP zero point and H$\zeta$ strengths.

\end{enumerate}

Our comparison of the stellar populations in the  RXJ0848.6+4453 galaxies
with that of more massive clusters at similar redshift and with less massive
clusters at higher redshift raise the possibility that the quenching
of star formation in the cluster galaxies depend on the cluster properties.
The quenching may either happen earlier or faster in the more massive clusters.
Detailed spectroscopic investigations of additional massive clusters at
$z=1-2$ are required to shed further light on this issue.

\vspace{0.5cm}

Acknowledgments:
Karl Gebhardt is thanked for making his kinematics software available.
Masayuki Tanaka is thanked for alerting us to the most recent mass estimate for PKS1138--262.
We thank the anonymous referee for constructive suggestions that helped improve
this paper.
The Gemini TACs and the former Director Fred Chaffee are thanked for generous time allocations
to carry out these observations.

Based on observations obtained at the Gemini Observatory, which is operated by the
Association of Universities for Research in Astronomy, Inc., under a cooperative agreement
with the NSF on behalf of the Gemini partnership: the National Science Foundation (United
States), the National Research Council (Canada), CONICYT (Chile), the Australian Research Council
(Australia), Minist\'{e}rio da Ci\^{e}ncia e Tecnologia (Brazil) 
and Ministerio de Ciencia, Tecnolog\'{i}a e Innovaci\'{o}n Productiva  (Argentina)

The data presented in this paper originate from the following Gemini programs:
GN-2011B-DD-3, GN-2011B-DD-5, and GN-2013A-Q-65.
In part, based on observations made with the NASA/ESA Hubble Space Telescope, 
obtained from the data archive at the Space Telescope Science Institute. 

I.J.\ acknowledge support from grant HST-AR-13255.01 from STScI.
STScI is operated by the Association of Universities for Research in Astronomy, 
Inc. under NASA contract NAS 5-26555.


\appendix

\begin{figure}
\begin{center}
\epsfxsize 8.5cm
\epsfbox{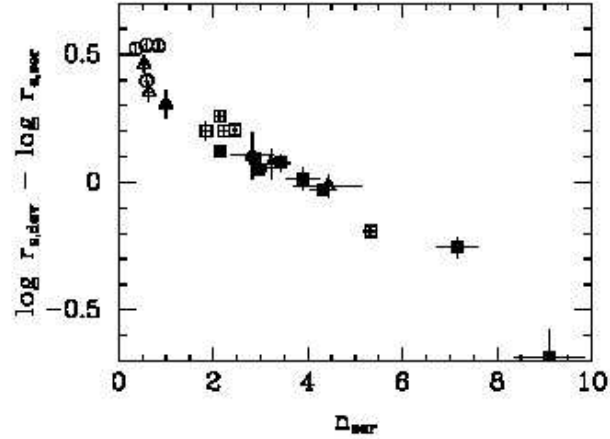}
\end{center}
\caption[]{
The difference $\log r_{\rm e,dev}-\log r_{\rm e,ser}$ versus the S\'{e}rsic index, $n_{\rm ser}$.
Solid squares -- bulge-dominated galaxies with EW[\ion{O}{2}] $\le 5${\AA}; 
open squares --  bulge-dominated galaxies with EW[\ion{O}{2}] $> 5${\AA}; 
circles - disk-dominated galaxies;
triangles - galaxies with S/N $<10$ in the spectroscopic observations and therefore excluded 
from the analysis.
As expected the figure shows a correlation, with the effective radius from the S\'{e}rsic fits
being larger than those from the fits with $r^{1/4}$-profiles if $n_{\rm ser}> 4$.
\label{fig-dlre_nser} }
\end{figure}

\begin{figure}
\begin{center}
\epsfxsize 8.5cm
\epsfbox{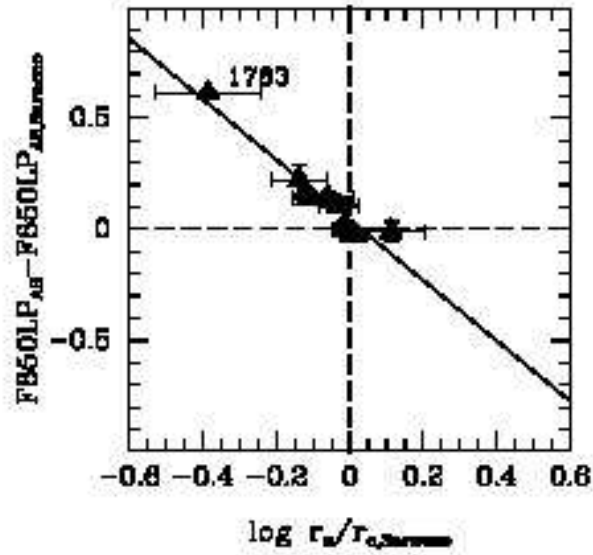}
\end{center}
\caption[]{
Comparison of our {\it HST} photometry with that of Saracco et al.\ (2014) for the eight galaxies in common,
show as the difference in the logarithm of the effective radii versus the difference in total magnitudes.
The errors in the two parameters are strongly correlated.
Solid line -- best fit relation. The zero point difference between our 
magnitudes and those of Saracco et al.\ (2014) as $0.04\pm 0.03$ mag.
\label{fig-phothst_comp} }
\end{figure}

\section{Photometry from {\it HST}/ACS \label{SEC-IMAGING}}

Table \ref{tab-photRXJ0848HST} lists the photometric parameters for the spectroscopic sample as derived
from the {\it HST}/ACS observations in F850LP and F775W. Only the F850LP images were processed to derive
2-dimensional surface photometry using GALFIT (Peng et al.\ 2002). F775W was used only for the color determinations.
The effective radii in Table \ref{tab-photRXJ0848HST} are derived from the semi-major and -minor axes as 
$r_{\rm e} = (a_{\rm e}\,b_{\rm e})^{1/2}$. The difference between the effective radii from fits with an
$r^{1/4}$-profile and a S\'{e}rsic profile should not be interpreted as the uncertainty. As expected
the difference is correlated with the S\'{e}rsic index, $n_{\rm ser}$, see Figure \ref{fig-dlre_nser}.

Figure \ref{fig-phothst_comp} shows our photometric parameters compared with those from Saracco et al.\ (2014) 
for the eight galaxies in common. We have offset the $z_{\rm 850,Vega}$ magnitudes from Saracco et al.\ (2014)
to AB magnitudes using the synthetic zero points from Sirianni et al.\ (2005).
As the errors on total magnitudes and effective radii are strongly correlated, we show the differences
in the logarithm of the effective radii (in kpc) versus the difference in the total $z_{\rm 850}$ magnitudes.
The solid line shows the best fit relation. From zero point of this fit we can derive the magnitude
difference between our magnitudes and those of Saracco et al.\ (2014) as $0.04\pm 0.03$ mag. This very
small difference is most likely due to the difference in the choice of point-spread-function (PSF).
Saracco et al.\ uses a PSF constructed from stars in the field, while we use Tiny Tim (Krist 1995) model PSFs (see
Chiboucas et al.\ 2009).

We have derived total magnitudes in the rest frame B-band from the observed $z_{\rm 850}$ magnitudes
and colors, using calibrations established based on
Bruzual \& Charlot (2003) stellar population models spanning the observed color range,
see J\o rgensen et al.\ (2005) for details. 
We use the aperture color $(i_{775} - z_{850} )$ in the calibration, and calibrate all total 
magnitudes to the rest frame B-band using this color. This ignores any effects of color gradients, but 
is preferable because the aperture color has lower uncertainty than the total color.
For the cluster redshift the calibration is
\begin{equation}
B_{\rm rest}=  z_{\rm 850} +1.390-1.144 (i_{775} - z_{850}) 
\end{equation}
The large color term is due to the fact that at the cluster redshift the F850LP filter spans the 
4000{\AA} break.
The distance modulus for our adopted cosmology is $DM(z)$ = 44.74.
The absolute B-band magnitude, $M_{\rm B}$, is then derived as
\begin{equation}
M_{\rm B} = B_{\rm rest} - DM(z).
\end{equation}
Techniques for how to calibrate to a ``fixed-frame'' photometric system are described in
detail by Blanton et al.\ (2003).

\begin{deluxetable*}{rrrr rrrr rrrr}
\tablecaption{RXJ0848.6+4453: Photometric Parameters from {\it HST}/ACS data \label{tab-photRXJ0848HST} }
\tablewidth{0pc}
\tablehead{
\colhead{ID} & \colhead{RA (J2000)} & \colhead{DEC (J2000)\tablenotemark{a}} & \colhead{$m_{\rm tot,SEx}$} & \colhead{$(i_{775}-z_{850})$}&  
\colhead{$m_{\rm tot,dev}$} & \colhead{$\log r_{\rm e,dev}$} & \colhead{$m_{\rm tot,ser}$} & \colhead{$\log r_{\rm e,ser}$} 
& \colhead{$n_{\rm ser}$} & \colhead{PA} & \colhead{$\epsilon$} \\
\colhead{(1)} & \colhead{(2)} & \colhead{(3)} & \colhead{(4)} & \colhead{(5)} & \colhead{(6)} & \colhead{(7)} 
& \colhead{(8)} & \colhead{(9)} & \colhead{(10)} & \colhead{(11)} & \colhead{(12)}
}
\startdata
  240&   8 48 48.20&   44 52 08.9&  23.00&  1.068&  22.85& -0.807&  22.56& -0.553&    7.2&   28.3&   0.19\\
  336&   8 48 47.33&   44 51 47.5&  24.53&  0.950&  24.52& -0.596&  24.41& -0.713&    1.7&  -29.4&   0.70\\
  361&   8 48 49.26&   44 54 17.3&  22.29&  0.903&  21.97& -0.215&  21.94& -0.189&    4.2&   72.9&   0.13\\
  438&   8 48 46.12&   44 51 23.8&  24.46&  0.527&  23.74& -0.076&  24.35& -0.499&    1.4&   81.7&   0.11\\
  634&   8 48 46.59&   44 53 17.0&  24.14&  0.556&  24.03& -0.582&  24.31& -0.776&    0.9&   43.3&   0.64\\
  654&   8 48 46.74&   44 53 36.6&  24.71&  0.535&  24.35& -0.424&  24.48& -0.528&    2.8&   -0.5&   0.73\\
  661&   8 48 47.17&   44 54 28.7&  21.20&  0.748&  20.76& -0.114&  21.05& -0.330&    2.2&  -16.6&   0.36\\
  722&   8 48 44.91&   44 52 26.6&  23.49&  0.710&  22.66&  0.097&  23.50& -0.466&    0.6&  -46.5&   0.65\\
  807&   8 48 44.27&   44 52 22.8&  24.18&  0.625&  23.79& -0.506&  24.05& -0.708&    1.9&   12.4&   0.42\\
  887&   8 48 45.59&   44 54 30.7&  24.25&  0.018&  23.60& -0.193&  23.96& -0.453&    2.1&  -28.6&   0.60\\
 1044&   8 48 43.29&   44 53 48.4&  22.84&  0.550&  22.20& -0.157&  22.85& -0.602&    0.6&   51.5&   0.67\\
 1045&   8 48 42.94&   44 53 06.6&  24.37&  0.042&  23.78& -0.363&  24.32& -0.729&    0.7&  -60.1&   0.51\\
 1123&   8 48 41.81&   44 52 45.4&  24.10&  0.658&  23.43& -0.150&  24.11& -0.613&    0.5&    9.0&   0.29\\
 1173&   8 48 42.73&   44 54 17.3&  23.49&  0.549&  22.81& -0.181&  23.44& -0.615&    0.4&  -59.1&   0.39\\
 1177&   8 48 40.36&   44 51 56.3&  23.78&  0.552&  23.14& -0.143&  23.74& -0.575&    0.6&  -54.2&   0.38\\
 1264&   8 48 39.66&   44 51 49.0&  23.92&  0.827&  23.68& -0.712&  23.70& -0.724&    3.9&  -85.3&   0.05\\
 1276&   8 48 40.07&   44 52 50.4&  22.76&  0.296&  22.18& -0.019&  22.71& -0.406&    0.7&   71.2&   0.84\\
 1352&   8 48 39.28&   44 52 10.8&  24.78&  0.569&  24.32& -0.266&  24.67& -0.535&    1.4&  -59.8&   0.81\\
 1362&   8 48 38.64&   44 52 12.5&  24.38&  0.728&  22.03&  0.143&  22.34& -0.114&    2.1&   50.0&   0.58\\
 1517&   8 48 39.36&   44 53 44.8&  23.05&  0.804&  22.87& -0.412&  22.86& -0.413&    3.8&   18.7&   0.49\\
 1533&   8 48 40.81&   44 55 11.4&  24.59&  0.682&  24.01& -0.401&  24.55& -0.753&    0.6&   22.1&   0.63\\
 1644&   8 48 37.96&   44 54 02.4&  23.19&  0.916&  21.72&  0.392&  22.31& -0.132&    0.4&   71.3&   0.35\\
 1698&   8 48 37.45&   44 53 26.7&  24.45&  0.557&  24.07& -0.689&  24.37& -0.893&    1.5&   10.4&   0.57\\
 1748&   8 48 37.07&   44 53 33.9&  23.09&  0.950&  22.81& -0.566&  22.93& -0.658&    2.9&   24.1&   0.39\\
 1763&   8 48 35.97&   44 53 36.0&  21.70&  1.043&  21.32&  0.120&  21.56& -0.085&    2.4&  -83.6&   0.24\\
 1809&   8 48 36.96&   44 53 56.2&  24.45&  0.975&  23.99& -0.256&  24.36& -0.562&    1.0&  -72.4&   0.50\\
 1888&   8 48 36.16&   44 54 17.2&  22.12&  0.949&  22.00& -0.331&  21.76& -0.139&    5.3&  -25.4&   0.24\\
 2015&   8 48 34.05&   44 53 02.4&  24.43&  0.909&  24.24& -1.125&  24.21& -1.110&    4.4&   57.2&   0.13\\
 2063&   8 48 34.07&   44 53 32.2&  23.17&  0.803&  22.75& -0.361&  23.01& -0.563&    2.2&   40.2&   0.28\\
 2111&   8 48 33.57&   44 53 44.0&  23.10&  0.666&  22.87& -0.359&  22.10&  0.324&    9.1&  -71.6&   0.34\\
 2138&   8 48 33.31&   44 53 27.0&  24.06&  0.529&  23.54& -0.256&  24.14& -0.662&    0.8&   86.7&   0.52\\
 2336&   8 48 26.67&   44 53 19.0&  21.71&  0.647&  21.21& -0.005&  21.61& -0.304&    1.6&   16.5&   0.59\\
 2342&   8 48 27.94&   44 54 51.0&  24.89&  0.542&  24.86& -0.644&  24.88& -0.832&    0.6&   47.3&   0.83\\
 2369&   8 48 29.39&   44 54 41.8&  24.30&  1.015&  22.94& -0.014&  23.50& -0.412&    0.6&  -36.4&   0.82\\
 2417&   8 48 28.24&   44 54 22.1&  23.31&  0.790&  22.53&  0.063&  23.35& -0.471&    0.8&  -41.0&   0.51\\
 2450&   8 48 28.59&   44 54 41.5&  24.03&  0.970&  23.83& -1.046&  23.83& -1.046&    4.0&  -61.3&   0.31\\
 2497&   8 48 26.93&   44 54 30.2&  24.17&  0.607&  \nodata&  \nodata&  23.87& -0.641&    2.8&  -86.2&   0.35\\
 2600&   8 48 33.00&   44 55 11.8&  23.14&  0.945&  22.92& -0.600&  22.67& -0.386&    6.2&   16.3&   0.32\\
 2624&   8 48 28.69&   44 53 03.0&  24.31&  0.828&  24.11& -0.736&  24.06& -0.698&    4.5&   10.0&   0.22\\
 2651&   8 48 30.56&   44 54 54.9&  23.44&  0.778&  22.12& -0.013&  22.97& -0.582&    0.3&   -4.2&   0.35\\
 2702&   8 48 29.68&   44 53 23.9&  24.33&  0.959&  24.03& -0.627&  24.12& -0.698&    3.2&  -18.3&   0.16\\
 2735&   8 48 27.43&   44 55 28.4&  23.32&  0.944&  23.16& -0.899&  23.12& -0.871&    4.3&  -62.1&   0.49\\
 2772&   8 48 30.79&   44 53 34.8&  23.03&  0.568&  22.27&  0.003&  23.07& -0.535&    0.6&  -76.9&   0.54\\
 2943&   8 48 32.76&   44 54 07.1&  23.42&  0.862&  23.15& -0.581&  23.25& -0.659&    3.4&   78.5&   0.36\\
 2989&   8 48 24.42&   44 56 09.0&  23.16&  0.932&  22.90& -0.816&  22.98& -0.863&    3.0&   27.5&   0.59\\
 3030&   8 48 32.74&   44 54 45.4&  22.71&  1.075&  22.12&  0.093&  22.50& -0.209&    1.8&   47.5&   0.33\\
 3074&   8 48 21.17&   44 54 33.1&  22.81&  1.003&  22.66& -0.723&  22.53& -0.610&    5.5&  -21.0&   0.52\\
 3090&   8 48 32.97&   44 53 46.6&  22.49&  0.971&  22.16& -0.675&  22.34& -0.798&    2.2&   80.0&   0.51\\
 3144&   8 48 21.37&   44 54 32.6&  24.12&  0.501&  22.73& -0.307&  23.18& -0.615&    0.8&  -78.3&   0.64\\
 3384&   8 48 26.54&   44 55 18.5&  23.12&  0.733&  22.14&  0.299&  22.56&  0.004&    2.5&  -49.6&   0.20\\
 3426&   8 48 25.51&   44 55 46.4&  23.88&  0.829&  23.58& -0.614&  \nodata&  \nodata&   $>4$\tablenotemark{a}&   72.3&   0.47\\
 3461&   8 48 24.70&   44 54 13.8&  22.91&  0.975&  22.40& -0.110&  22.95& -0.468&    0.9&   13.0&   0.12\\
 3505&   8 48 22.87&   44 53 20.5&  23.41&  0.851&  22.32& -0.293&  22.95& -0.696&    1.0&   17.0&   0.24\\
 9001&   8 48 19.43&   44 53 46.5&  24.32&  0.508&  24.36& -0.523&  24.66& -0.709&    1.4&  -44.6&   0.77\\
\enddata
\tablenotetext{a}{No reliable S\'{e}rsic fit can be derived due to scattered light from a nearby source.}
\tablecomments{Col.\ (1) Galaxy ID; 
col.\ (2) and (3) Positions consistent with USNO (Monet et al.\ 1998), with an rms scatter of $\approx 0.5$ arcsec.
col.\ (4) total F850LP magnitude from SExtractor;
col.\ (5) Aperture color within an aperture with radius 0.5 arcsec;
col.\ (6) total magnitude from fit with $r^{1/4}$ profile;
col.\ (7) logarithm of the effective radius in arcsec from fit with $r^{1/4}$ profile; 
col.\ (8), (9) total magnitude and logarithm of the effective radius, from fit with a S\'{e}rsic profile;
col.\ (10) S\'{e}rsic index, the typical uncertainty is 0.1, while the uncertainties for ID 2015 and 2111 are 0.7; col.\ (11) position angle of major axis measured from North through East; col.\ (12) ellipticity.
}
\end{deluxetable*}

\section{Spectroscopic Data \label{SEC-SPECTROSCOPY}}

\subsection{Observations and reductions}

The spectroscopic reductions were done using the same techniques as described
for the nod-and-shuffle data used in J\o rgensen \& Chiboucas (2013). 
The only difference is that size and behavior of the  charge diffusion effect (CDE) of the the GMOS-N E2V 
DD CCDs compared to that of the original GMOS-N E2V CCDs.
Briefly, the CDE becomes stronger at longer wavelengths and is seen as charge diffused from
each individual pixel into all neighboring pixels, see also Abraham et al.\ (2004) for 
a description and J\o rgensen \& Chiboucas (2013) for a figure that shows how the 
CDE affects spectroscopic data.  As explained in that paper, in the absence of a 
correction the CDE leads to over-subtraction of the sky signal at long wavelengths. 

We attempted to use the same technique to correct for the CDE for the E2V DD CCDs as used
in J\o rgensen \& Chiboucas (2013). However, it turns out that the effect for these CCDs not 
only depends on the wavelength but also on the physical location on the array. 
We therefore obtained custom calibrations to enable the correction. These calibrations consist
of quartz-halogen flat fields taken through the MOS masks. 
The flat fields are shuffled on the detector the same way as the science observations are obtained. 
However, in order to avoid saturating the array only one shuffle cycle was used for the shuffled 
flat fields.
At the same time we obtained conventional quartz-halogen flat fields. 
Using the two types of flat fields together makes in possible to derive a correction image that can
be used to correct the flat fields taken with the science observations at night, such that after
the correction the flat fields will include the multiplicative effect of the CDE.
Due to small flexure effect as well as the fact that the science data are taken with deliberate 
small offsets in the detector translation stage between exposures, flat fielding for pixel-to-pixel
variations should be done with flat fields taken with the science data, rather than simply use the
shuffled flat fields obtained after the fact.

\begin{figure*}
\epsfxsize 17.5cm
\epsfbox{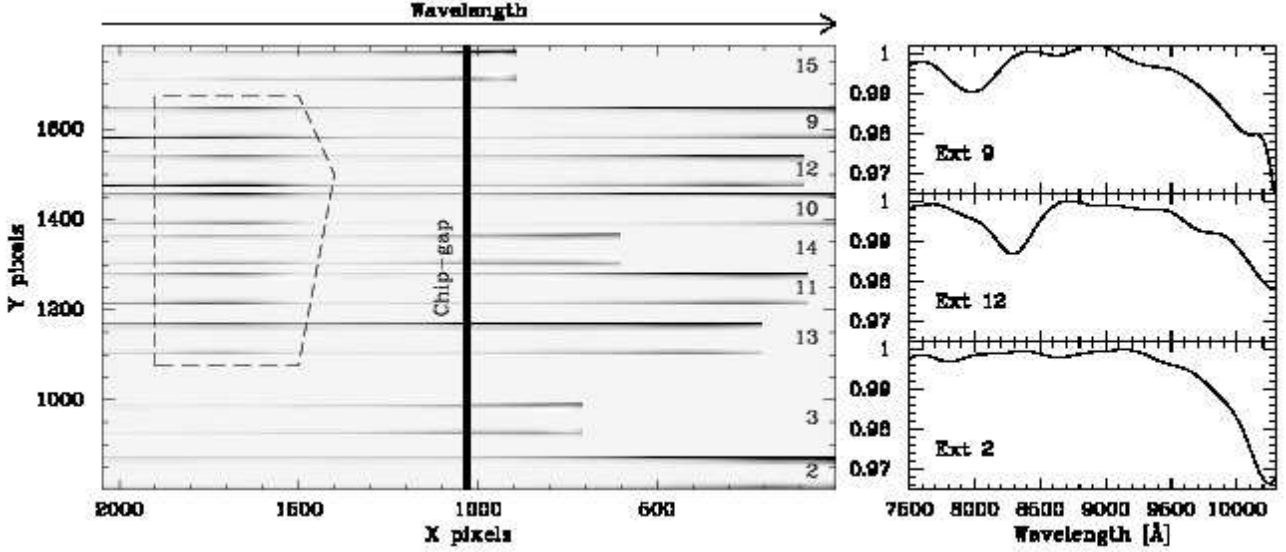}
\caption[]{
Left panel: Subimage of the CDE correction image. The image is scaled from 0.92 to 1.003. 
White areas either have insignificant correction (values of $\approx 1$) or 
are between the slitlets. Grey-to-black shows larger effects. 
Black vertical area -- chip gap between CCD1 and CCD2.
Each slitlet can be seen on the image as an area near symmetric in the Y-direction as the
correction in the upper and lower part of the shuffled image of the slitlet is expected to
be identical. The subimage contains nine slitlets, labeled with their extension number.
The dashed line marks the approximate area on CCD2 where the CDE varies with spatial 
location on the CCD, in addition to the variation with wavelength, see text.
Right panels: The CDE for three example extensions (2, 9 and 12) as it affects the science data.
Extensions 9 and 12 are affected by the area on CCD2 where the CDE varies with spatial  
location on the CCD.
\label{fig-cde_grey} }
\end{figure*}

The determination of the correction image is done with the following steps:
\begin{enumerate}
\item
Bias correct of both shuffled and conventional flat field.
\item
Fit both flat fields with cubic splines row-by-row and detector-by-detector.
\item
Normalize the smooth fits for each slitlet with the fit in the central row of each slitlet. 
This is also done detector-by-detector.
\item
For both flat fields, mosaic the images from the three arrays in GMOS-N.
\item
Cut each flat field into slitlets.
\item
Ratio the conventional flat field with the shuffled flat field in original position. 
This will show the CDE affecting the lower part of the lower image of the slitlets.
\item
Repeat the steps mosaic, cut and ratio, but with the conventional flat field offset to match the 
upper image of the flat field in the shuffled flat field. 
The resulting ratio image will show the CDE affecting the upper part of the upper image of the slitlets
\item
Fit the effect row-by-row with a cubic spline in order to suppress pixel-to-pixel noise in the 
correction image.
\item
Reformat the correction image to match the flat fields, which have six extensions matching 
the six read-out amplifiers on array.
\end{enumerate}
The calibration was repeated for each of the two wavelength setting used for the science observations,
and for each of the two MOS masks. Thus, in total four calibration images were used.
We repeated the determination several times over a period of a few weeks to ensure that the 
calibration did not change in time. No significant time dependence was found.
Figure \ref{fig-cde_grey} shows a sub-image of the 2-dimensional correction image together with examples 
of the size of the correction as it affects the science data. 
The size of the correction depends on the geometry of 
the slitlets, the placement of the objects in the slitlets, and the extraction aperture. 
The correction effect shown on Figure \ref{fig-cde_grey} matches the configuration used for our data.
The CDE calibration is critical for MOS nod-and-shuffle observations using as short slits as done
here (2.5 arcsec) and attempting to obtain observations redwards of $\approx 8000${\AA}. 

Following the CDE correction, the remainder of the spectroscopic reduction steps are identical
to those described in J\o rgensen \& Chiboucas (2013).
The result of the spectroscopic reductions are 1-dimensional extracted spectra calibrated to
a relative flux scale. The spectra were median filtered and resampled to just better than
critical sampling, cf.\ J\o rgensen \& Chiboucas. The instrumental resolution was derived
from sky spectra processed exactly the same way as the science spectra.  
The instrumental resolution for all slitlets is between
2.90 and 3.14 {\AA} measured as sigma in a Gaussian fit to the sky lines.

\begin{figure*}
\epsfxsize 17.5cm
\epsfbox{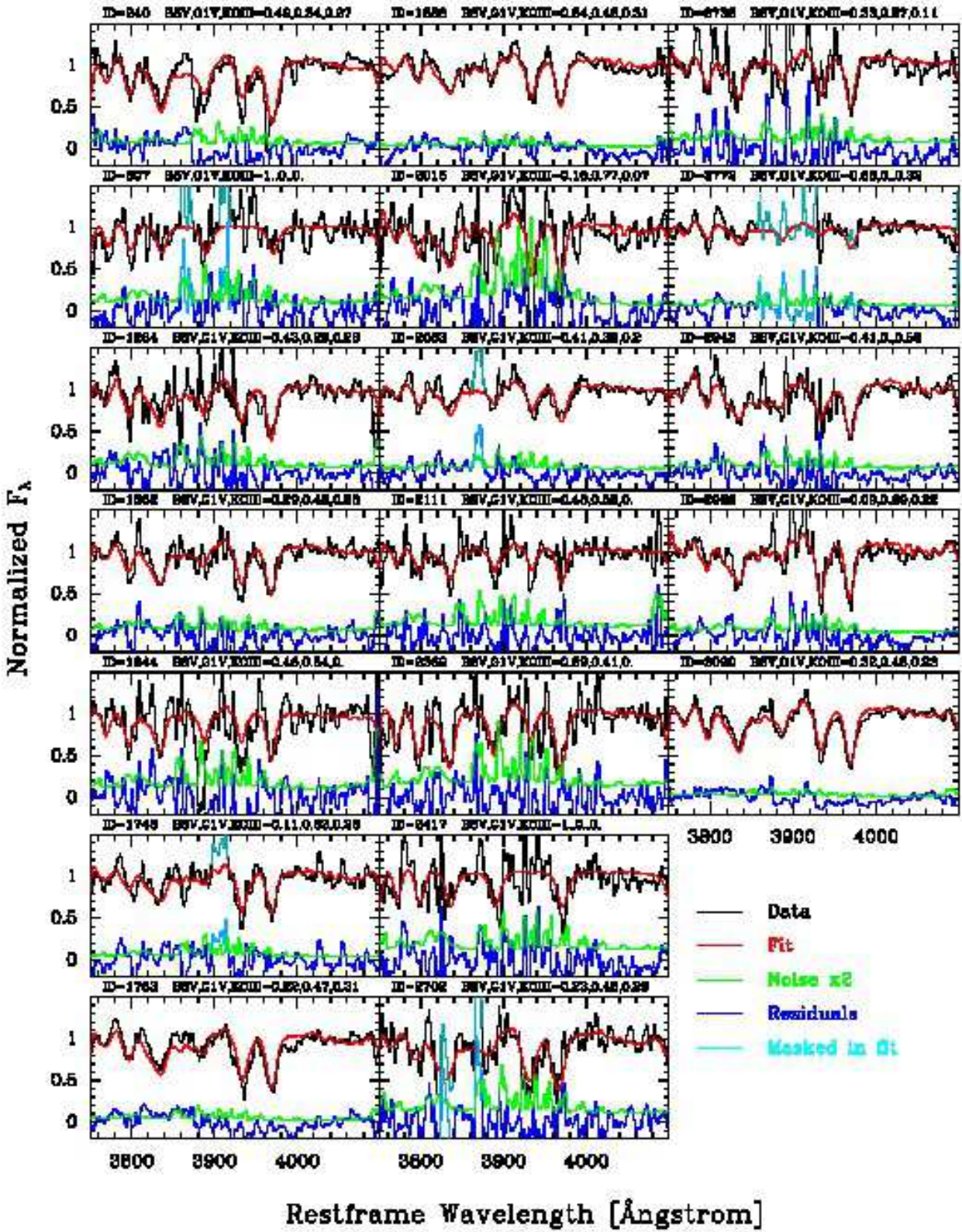}
\caption[]{
Summary of the kinematics fitting for the members of RXJ0848.6+4453 for which determination
of the velocity dispersion was possible.
Black -- normalized spectra ; red - best fit; green -- noise times two, normalized the same way as
the spectra; blue - residuals from the best fit; cyan -- wavelength regions excluded from the 
fits due to strong sky subtraction residuals.
The figure shows the general quality of the fits and demonstrates that the use of the three
template stars adequately spans the stellar populations present in the sample.
\label{fig-kinfit} }
\end{figure*}

\begin{figure*}
\begin{center}
\epsfxsize 8.5cm
\epsfbox{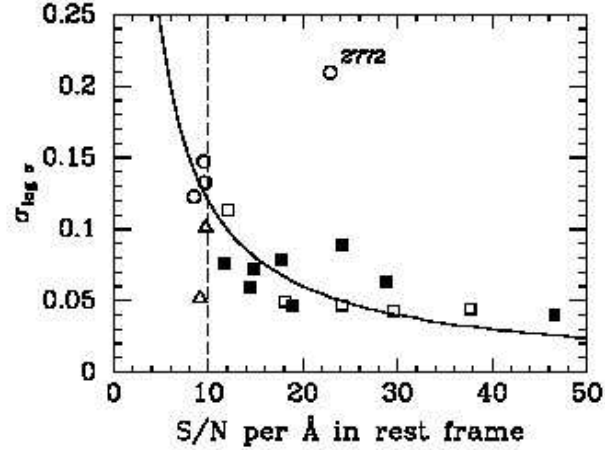}
\end{center}
\caption[]{
The uncertainty on the velocity dispersions as derived using a bootstrap method versus the 
S/N per {\AA} in the rest frame of the galaxies. Solid squares -- bulge-dominated galaxies 
with EW[\ion{O}{2}] $\le 5${\AA}; open squares --  bulge-dominated galaxies 
with EW[\ion{O}{2}] $> 5${\AA}; triangles - bulge-dominated galaxies with S/N $<10$; 
circles -- disk-dominated galaxies. Galaxy ID 2772 marked on the figure is an irregular galaxy, 
see imaging on Figure \ref{fig-specRXJ0848}, for which masking of residuals from the sky subtraction
affected the available wavelength range significantly, cf.\ Figure \ref{fig-kinfit}.
Solid line -- approximate relation between S/N and $\sigma _{\log \sigma}$: 
$\sigma _{\log \sigma} \approx 1.2 \times {\rm S/N^{-1}}$.
Dashed vertical line -- S/N=10. Galaxies below this S/N cutoff are excluded from the determination
of relations and zero points.
\label{fig-SNelsigma} }
\end{figure*}

\begin{figure*}
\epsfxsize 17cm
\epsfbox{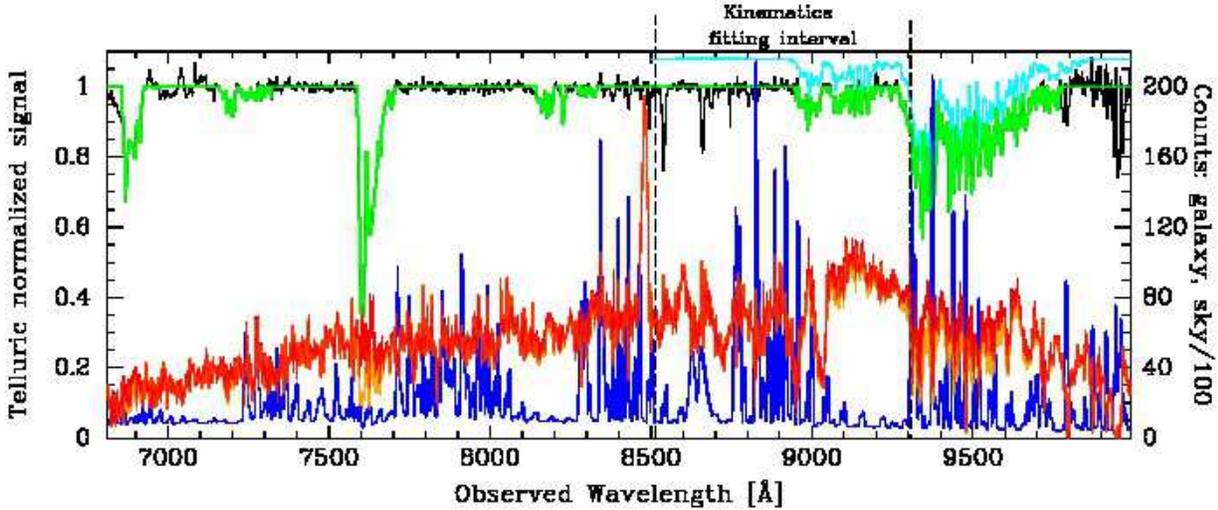}
\caption[]{
Telluric correction and typical sky spectrum for illustration shown together with the spectrum 
of the brightest cluster member observed.
Black - combined spectra of the two blue stars in the masks. 
Cyan -- ATRAN model (Lord 1992) for water vapor of 2 mm and elevation of 45 degrees at the altitude of Gemini North. 
The model has been resampled and convolved to match the spectral sampling and instrumental resolution of our data.
The model is offset 0.08 from the data for clarity.
Green -- adopted telluric correction.
Orange -- uncalibrated spectrum (in counts) of the brightest cluster member observed ID 1763.
Red -- spectrum (in counts) of ID 1763 after telluric correction.
Blue -- sky spectrum matching ID 1763, divided by 100.
See text for discussion of the figure.
\label{fig-tell_abs} }
\end{figure*}

\begin{figure*}
\epsfxsize 17.5cm
\epsfbox{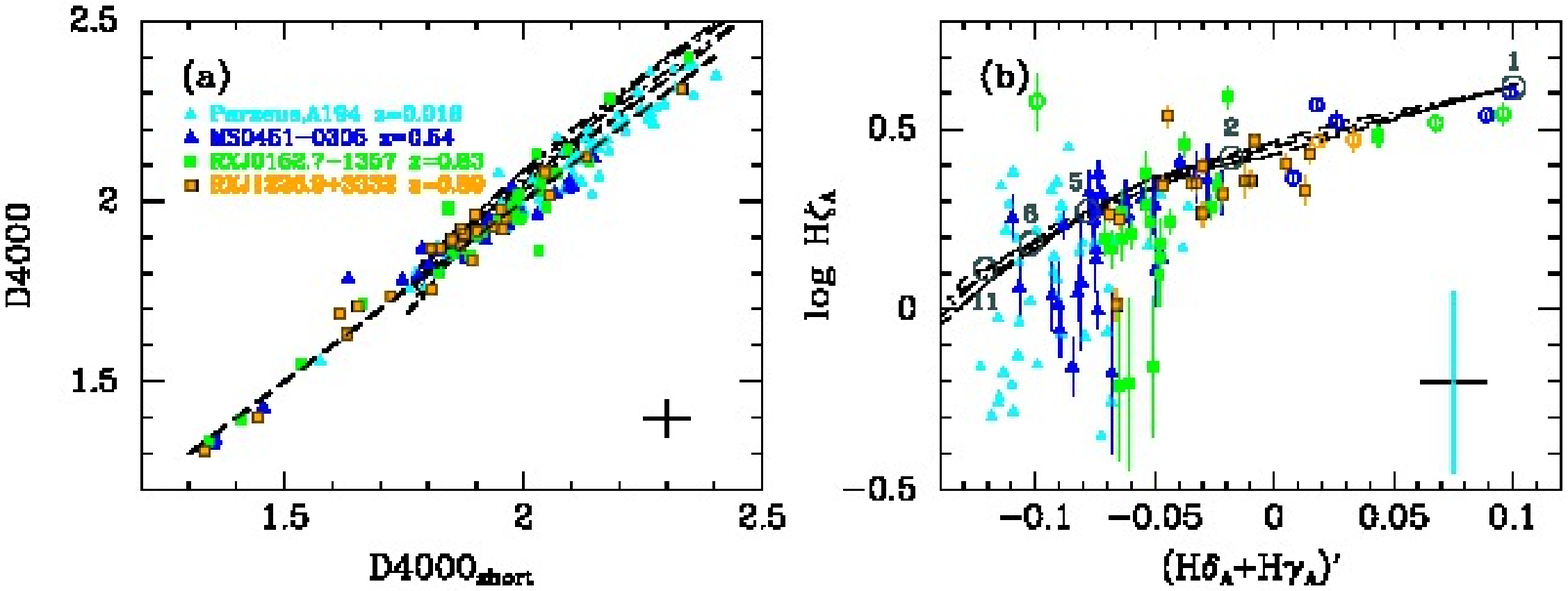}
\caption[]{
(a) Calibration of D4000$_{\rm short}$ to consistency with D4000. Typical measurement uncertainties 
are shown based on repeat measurements (see J\o rgensen \& Chiboucas 2013 for details).
Cyan ponts -- Perseus, A194; Blue points -- MS0451--0305; green points -- RXJ0152.7--1357; 
orange points -- RXJ1226.9+3332.
Thick dashed line -- one-to-one relation, scatter relative to this relation is 0.046 in D4000. 
Thin dashed and dotted lines -- predictions based on deriving the indices from the model SEDs 
from Maraston \& Str\"{o}mb\"{a}ck (2011).
(b) Higher order Balmer line indices versus each other. 
$({\rm H\delta _A + H\gamma _A})' \equiv -2.5~\log \left ( 1.-({\rm H\delta _A + H\gamma _A})/(43.75+38.75) \right )$, see Kuntschner (2000).
The typical measurement uncertainty on $({\rm H\delta _A + H\gamma _A})'$
is shown in the lower right of the panel. The adopted value of 0.014 is based on the internal comparison 
of repeat measurements presented in J\o rgensen \& Chiboucas (2013).
For H$\zeta _{\rm A}$ we show individual uncertainties for the $z=0.5-1$
clusters as these vary significantly as a function of the strength of the line. 
For the low redshift sample the median uncertainty on $\log {\rm H}\zeta _{\rm A}$ is 0.26, as shown 
on the figure.
Symbols as on panel (a), except galaxies with  S\'{e}rsic index less than 1.5 are shown as open circles.
Thin dashed and dotted lines -- predictions based on deriving the indices from the model SEDs
from Maraston \& Str\"{o}mb\"{a}ck. Open circles on the model lines correspond to the models with solar [M/H] and
ages in Gyr as labeled, see text for discussion. 
\label{fig-d4000Hindex} }
\end{figure*}

\subsection{Spectroscopic parameters}

The calibrated spectra were fit with stellar templates as described in J\o rgensen \& Chiboucas (2013).
This results in determination of the redshifts and the velocity dispersions.
The fits were performed with the kinematics fitting software made available by Karl Gebhardt, see
Gebhardt et al.\ (2000, 2003) for a description of the software.
Spectra of member galaxies were fit in the wavelength range 3750--4100{\AA}. 
The software fits the spectra in pixel space. The template stars are convolved to the 
instrumental resolution of the science spectra. Individual values for each slit is used as these
are not identical. As in J\o rgensen \& Chiboucas (2013) we determine the instrumental resolution
from stacked sky spectra processed in the same way as the science spectra.
The software determines the line-of-sight-velocity-distribution (LOSVD) from the science spectra
and then the velocity dispersion is derived from the LOSVD through both a Gauss-Hermite polynomial fit
and a Gaussian fit. For the 19 member galaxies with velocity dispersion determined the offset in 
log space between the two measurements is $-0.036$ with the Gauss-Hermite polynomial fits giving the 
smaller velocity dispersions.  
One may argue that using Gaussian fits is a suitable choice for the RXJ0848.6+4453 data due to 
the relatively short wavelength range and a median S/N of 18 per {\AA}ngstrom, making the uncertainties
on the higher order terms of the Gauss-Hermite polynomial fit large.
However, for consistency with previous work, we use the velocity dispersions 
derived from the Gauss-Hermite polynomial fits to the LOSVD.  

Because the software determines the fits in pixel space, it is straight forward to mask wavelength ranges 
not to be included in the fit. We used this to flag areas of strong residuals from the sky subtraction, 
and for ID 2772 also to mask Balmer emission lines in the fitted wavelength range.
None of the other galaxies have significant emission lines within the wavelength range 3750--4100{\AA}.
As in J\o rgensen \& Chiboucas, we use three template stars with spectral types K0III, G1V and B8V.
Figure \ref{fig-kinfit} shows the normalized spectra, fits and residuals in the wavelength
region covered by the fits. The purpose of the set of template stars is to span the stellar 
populations in the galaxies. From Figure \ref{fig-kinfit} we conclude that use of these three stars 
accomplishes this.  Further, we choose to use the same template stars as in our previous work 
to ensure consistency with our previous publications.
Aperture correction of the velocity dispersions were performed using the technique from 
J\o rgensen et al.\ (1995b).

The uncertainties, $\sigma _{\log \sigma}$, on the velocity dispersions are derived by the fitting 
software using a bootstrap method. In Figure \ref{fig-SNelsigma} we show the resulting uncertainties 
versus the S/N of the spectra. The uncertainties scale approximately with S/N as 
$\sigma _{\log \sigma} \approx 1.2 \times {\rm S/N^{-1}}$. This relation is shown on the figure. 

We have assessed possible systematic errors on the derived velocity dispersions resulting from
(1) incorrect instrumental resolution, (2) incorrect telluric correction, and (3) incorrect sky subtraction.

The instrumental resolution varies from slit to slit with the total range being approximately $\pm 4.5$\%.
The formal uncertainties on the derived instrumental resolutions as estimated from the rms scatter of
measurements of individual skylines in each slit is 1--3\%. We estimated the systematics that may be introduced
in the velocity dispersion measurements from incorrect determination of the instrumental resolution by
refitting all spectra adopting instrumental resolutions of 10\% larger and 10\% smaller than measured.
The median offsets relative to the velocity dispersions listed in Table \ref{tab-RXJ0848kin} are less than 0.005
in $\log \sigma$. As this is a factor five less than the intrinsic consistency of 0.026 our low redshift sample 
(J\o rgensen et al.\ 2005), we conclude that systematic errors from incorrect determination of
the instrumental resolution do not affect our results significantly.

The telluric correction was derived from spectra of two blue stars included in the masks. Thus, the
calibration spectra were obtained with identical airmass and water vapor as the science spectra and processed 
in the same way as the science spectra.
The combined spectrum of the two blue stars was normalized. We then cleaned the spectrum for 
noise and stellar features outside the known bands of atmospheric absorption. At wavelengths long than 8500 {\AA}
we used an ATRAN model spectrum of the atmospheric absorption (Lord 1992) to guide this cleaning process.
Figure \ref{fig-tell_abs} shows the normalized spectrum before cleaning, the ATRAN model, and the adopted 
telluric correction. The wavelength interval for the kinematics fitting is shown in the observed frame. 
The long wavelength limit of this fitting interval was set to avoid the strong telluric absorption bands
redwards of $\approx 9300$ {\AA}.
As the spectra used to determine the telluric correction are obtained simultaneously with the science,
there can be no mismatch in airmass or water vapor. Nevertheless, we tested the effect of the telluric
correction being systematically wrong with 10\% by correcting the spectra with the telluric correction
test spectrum derived from the measured telluric correction spectrum as 
\begin{equation}
{\rm telluric_{test}} = 1 - X (1-{\rm telluric_{measured}})
\end{equation}
with $X=0.9$ and 1.1, respectively. The spectra were then flux calibrated, resampled, normalized, processed
with the kinematics fitting software. The resulting velocity dispersions are in both cases consistent with
those presented in Table \ref{tab-RXJ0848kin} with median zero point offsets of $<0.003$ for the two cases
and rms scatter of 0.04 in $\log \sigma$ for the 19 member galaxies with measured velocity dispersions.

Finally we assessed the systematic effects on the velocity dispersions from a possible incorrect sky subtraction.
The data are obtained in nod-and-shuffle mode to limit such sky subtraction errors. Within the wavelength
fitting interval (3750--4100 {\AA} in the rest frame) even for the brightest cluster member observed, ID 1763, the 
signal from the galaxy is only 2.6\% of the average sky signal. Figure \ref{fig-tell_abs} illustrates
this by showing the spectrum of ID 1763 in counts together with the matching sky spectrum scaled down with
a factor 100.
We tested the effect of the sky subtraction being over- or under-subtracted with 0.1\%.
An error in the sky subtraction of this size cause visible systematic residuals 
in the resulting spectra within the wavelength region used for the kinematics fitting. 
We therefore view this as the upper limit on the possible systematic error in the sky subtraction.
The spectra were then processed through the remainder of the processing steps and velocity dispersions 
derived. Comparing the resulting velocity dispersions to those presented in 
Table \ref{tab-RXJ0848kin} for the 19 member galaxies with measured velocity dispersions
we find median offsets in $\log \sigma$ of $<0.015$ in the two cases.
The rms scatter in $\log \sigma$ for over- and under-subtraction of the sky is 0.13 and 0.05, respectively.
If limiting the comparison to the 13 bulge-dominated galaxies with S/N$\ge$10, the scatter
decreases to 0.08 and 0.03, respectively. These 13 galaxies are the galaxies included in our analysis
of the size and velocity dispersion evolution as well as the FP analysis.

We conclude that systematic errors on the the measured velocity dispersions due to the 
determination of the instrumental resolution, the adopted telluric correction or the sky 
subtraction do not affect our results significantly.

Our ability to determine line indices from the spectra are limited by the sky subtraction errors,
which in turn to a large extent are a result of the imperfect correction for the CDE.
While the spectra cover wavelengths to 4400{\AA} in the rest frame, the sky subtraction errors limit
the range for which the spectra can be used for determination of absorption line indices to 
$\approx 4075${\AA} in the rest frame (9250{\AA} observed).
We have derived the following indices CN3883, CaHK, D4000 and H$\zeta _{\rm A}$. 
We have adopted the bandpass definition for the higher order Balmer line H$\zeta$ from Nantais et al.\ (2013).
The index is named $\rm H6_{\rm A}$ in Nantais et al., while we opt to name it after the name of 
the Balmer line it is measuring.

The red bandpass of the original definition of the D4000 index (Gorgas et al.\ 1999) 
is in our spectra severely affected by the sky subtraction errors.
We have therefore defined a shorter red bandpass in order to be able to measure the 
equivalent of the D4000 index. We call this index D4000$_{\rm short}$ and define it as
\begin{equation}
{\rm D4000_{short}} = 1.2107 \,(200/75) \, 
\left( \int _{4000(1+z)}^{4075(1+z)} f_{\lambda} d\lambda \right) \, 
\left( \int _{3750(1+z)}^{3950(1+z)} f_{\lambda} d\lambda \right)^{-1} - 0.092
\label{eq-D4000}
\end{equation}
The definition ensures that on average ${\rm D4000_{short} = D4000}$.
The factor (200/75) reflects the difference in the length of the two passbands in the definition, 
as the original D4000 index has identical length passbands.
Figure \ref{fig-d4000Hindex}a shows the two indices versus each other for our sample of 
member galaxies in Perseus, A194, MS0451--0305, RXJ0152.7--1357, and RXJ1226.9+3332. 
The figure also shows the indices determined from
the model SEDs from Maraston \& Str\"{o}mb\"{a}ck (2011).
The rms scatter of the data points relative to the one-to-one relation is 0.046. 
In J\o rgensen \& Chiboucas we estimated that the measurement uncertainty on D4000 for our 
$z=0.5$ and $z=0.9$ data is 0.051 and 0.033, respectively. Thus, the internal scatter 
of the relation between D4000  and $\rm D4000_{short}$ is insignificant, and we will assume
that $\rm D4000_{short}$ has a similar uncertainty as that of D4000.

Our data for RXJ0848.6+4453 do not allow us to stack the frames in subsets for the full data set
in order to derive better estimate uncertainties on the spectroscopic parameters.
However, in the wavelength region 3750{\AA} to 4100{\AA} the median S/N of the data is similar
to that of our spectra for MS0451.6--0305 (J\o rgensen \& Chiboucas 2013). We therefore adopt the 
uncertainties for CN3883, log CaHK and D4000 from that paper (Table 19 in that paper).
The values are 0.035, 0.060 and 0.051 for CN3883, log CaHK and D4000, respectively. 
The uncertainty on  H$\zeta _{\rm A}$ depends too strongly on the strength of the index to be 
adequately represented by one value. Thus, we show individual error bars on figures
using this index, except for the low redshift sample for which we for clarity show the median 
uncertainty.
The low redshift sample spectra have fairly low S/N in the wavelength region of H$\zeta$. 
That combined with the weak strength of the line for these galaxies leads to a high relative uncertainty,
typically 0.26 on $\log {\rm H}\zeta _{\rm A}$. 
The line indices have been aperture corrected and corrected for velocity dispersion as described in 
J\o rgensen et al.\ (2005), see J\o rgensen \& Chiboucas (2013) for a discussion of the method
applied to intermediate redshift galaxies.
As for the lower order Balmer lines, we assume that H$\zeta _{\rm A}$ has no aperture correction.

As H$\zeta _{\rm A}$ is the only Balmer line index measurable from our RXJ0848.6+4453 data,
we have used the data for the three $z<1$ clusters as well as model SEDs from Maraston \& Str\"{o}mb\"{a}ck (2011)
to evaluate to what extent this index is suitable for estimating ages.
In Figure \ref{fig-d4000Hindex}b we show the indices for the high order Balmer lines versus each
other for the low redshift sample and three $z<1$ clusters. 
Model predictions based on the model SEDs from Maraston \& Str\"{o}mb\"{a}ck are overplotted.
H$\zeta _{\rm A}$ is useful as an indicator of very young stellar populations, ages of 2 Gyr or less. 
The ages of the older stellar populations are difficult to measure using H$\zeta _{\rm A}$
due to the uncertainties affecting our data.

The mix of the three template stars in our best fit models from the kinematics fitting as expected
correlate with the H$\zeta _{\rm A}$ and CaHK indices in the sense that the B8V fraction in the 
best fit model increases with increasing H$\zeta _{\rm A}$, while the sum of the G1V and K0III fractions
increases with increasing CaHK. However, because stellar population models already exist to assist
the interpretation of the index strengths and no such modeling exist for the template stars, we will
use only the indices in our analysis of the data.

For galaxies with detectable emission from  [\ion{O}{2}] we determined the equivalent width 
as well as the (relative) flux of the [\ion{O}{2}]$\lambda\lambda$3726,3729 doublet.
With an instrumental resolution of $\sigma \approx 3$\,{\AA} (FWHM $\approx 7$\,{\AA}), 
the doublet is not resolved in our spectra and we refer to it simply as the
``[\ion{O}{2}] line''. 

Tables \ref{tab-RXJ0848kin} and \ref{tab-RXJ0848line} list the results from the template fitting 
and the derived line strengths.

\section{Presentation of the imaging and spectra for RXJ0848.6+4453 \label{SEC-SPECIM}}

\begin{figure*}
\epsfxsize 17.5cm
\epsfbox{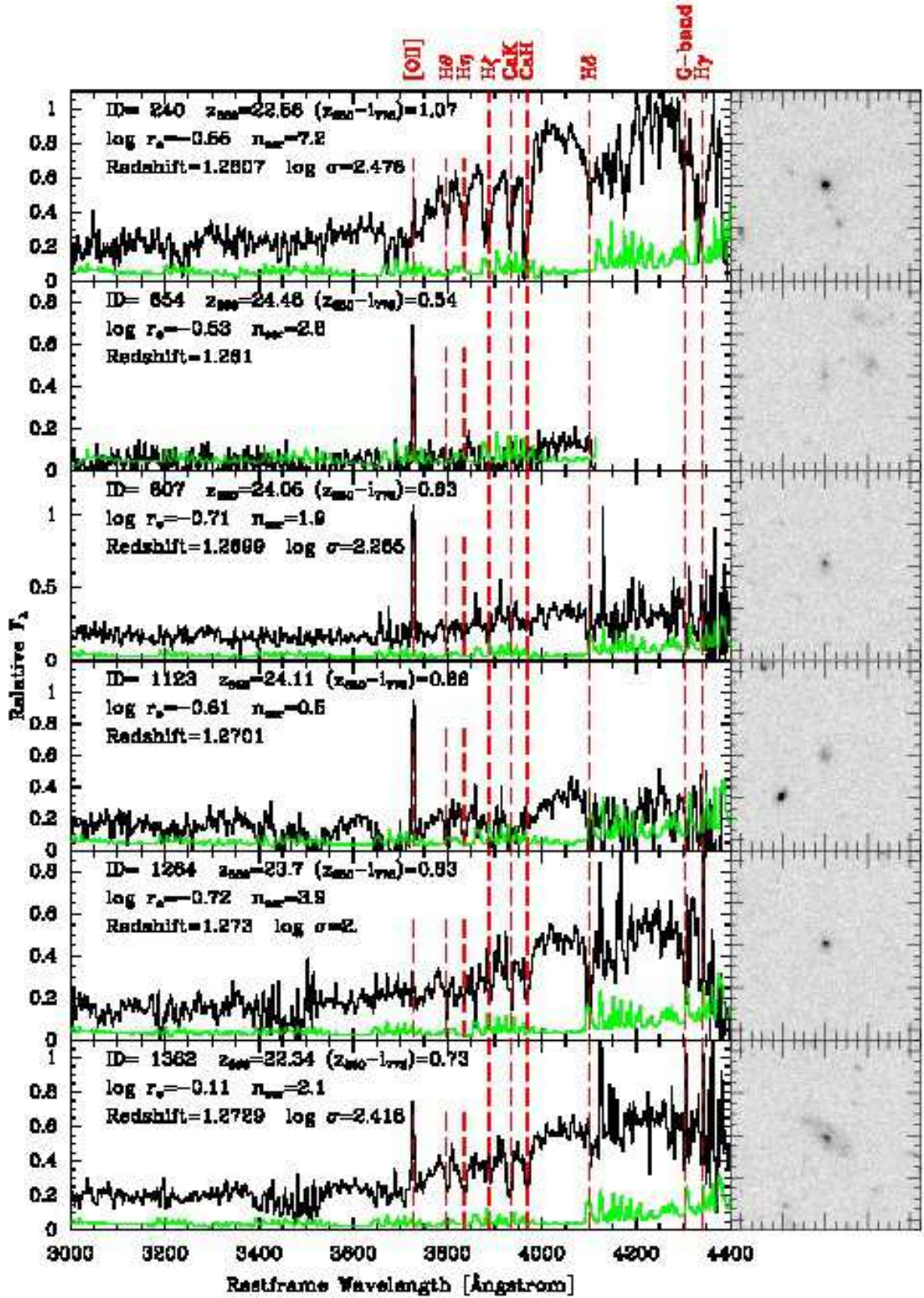}
\vspace{-1cm}
\caption[]{Spectra and grey scale images of the galaxies that are considered members of RXJ0848.6+4453.
On the spectra, black lines show the galaxy spectra, green lines show the random noise multiplied with two.
At the strong sky lines, the random noise underestimates the real noise due to systematic
errors in the sky subtraction. 
Some of the absorption lines are marked. The location of the emission line [\ion{O}{2}],
is also marked, though emission is only present in some of the galaxies. 
The spectra shown in this figure have been processed as described
in the text, including resampling to just better than critical sampling. 
The grey scale images are made from the {\it HST}/ACS images of the RXJ0848.6+4453 sample in the F850LP filter. 
Each image is 9 arcsec $\times $ 9 arcsec.
At the distance of RXJ0848.6+4453 this corresponds to 75 kpc $\times$ 75 kpc for our adopted cosmology. 
North is up, East to the left.
\label{fig-specRXJ0848} }
\end{figure*}

The spectra as well as stamp-sized images of the galaxies from the {\it HST}/ACS 
imaging of the cluster members are shown in Figure \ref{fig-specRXJ0848}.
The stamps cover the equivalent of 75 kpc $\times$ 75 kpc at the distance of the cluster.
The spectra used to create Figure \ref{fig-specRXJ0848} are available in the online journal.

\begin{deluxetable*}{rrrrrrrrrrr}
\tablecaption{RXJ0848.6+4453: Results from Template Fitting \label{tab-RXJ0848kin} }
\tablewidth{400pt}
\tabletypesize{\scriptsize}
\tablehead{
\colhead{ID} & \colhead{Redshift} & \colhead{Member\tablenotemark{a}} &  
\colhead{$\log \sigma$} & \colhead{$\log \sigma _{\rm cor}$\tablenotemark{b}} & 
\colhead{$\sigma _{\log \sigma}$} & \multicolumn{3}{c}{Template fractions} & \colhead{$\chi ^2$} & \colhead{S/N\tablenotemark{c}} \\
\colhead{}&\colhead{} &\colhead{} &\colhead{} &\colhead{} &\colhead{} & \colhead{B8V} & \colhead{G1V} & \colhead{K0III} & }
\startdata
  240& 1.2607&  1&  2.452&  2.476&  0.046&   0.49&   0.24&   0.27&    3.2&   18.9\\
  336& \nodata&  2&  \nodata&  \nodata&  \nodata&  \nodata&  \nodata&  \nodata&  \nodata&    5.3\\
  361& 1.3285&  0&  2.346&  2.370&  0.060&   0.00&   0.87&   0.13&    4.2&   23.2\\
  438& 1.3276&  0&  \nodata&  \nodata&  \nodata&  \nodata&  \nodata&  \nodata&  \nodata&    2.8\\
  634& 1.3255&  0&  \nodata&  \nodata&  \nodata&  \nodata&  \nodata&  \nodata&  \nodata&    5.1\\
  654& 1.2610&  1&  \nodata&  \nodata&  \nodata&  \nodata&  \nodata&  \nodata&  \nodata&    2.9\\
  661& 0.8455&  0&  2.216&  2.239&  0.059&   0.20&   0.80&   0.00&    7.2&   61.8\\
  722& 1.1374&  0&  \nodata&  \nodata&  \nodata&  \nodata&  \nodata&  \nodata&  \nodata&    8.5\\
  807& 1.2699&  1&  2.241&  2.265&  0.114&   1.00&   0.00&   0.00&    2.1&   12.1\\
  887& 0.7160&  0&  \nodata&  \nodata&  \nodata&  \nodata&  \nodata&  \nodata&  \nodata&    8.7\\
 1044& 0.5440&  0&  2.631&  2.651&  0.057&   0.25&   0.00&   0.75&    1.7&   17.0\\
 1045& 0.4328&  0&  \nodata&  \nodata&  \nodata&  \nodata&  \nodata&  \nodata&  \nodata&    8.9\\
 1123& 1.2701&  1&  \nodata&  \nodata&  \nodata&  \nodata&  \nodata&  \nodata&  \nodata&    6.8\\
 1173& 1.1162&  0&  \nodata&  \nodata&  \nodata&  \nodata&  \nodata&  \nodata&  \nodata&   12.4\\
 1177& 1.2346&  0&  2.205&  2.229&  0.187&   0.70&   0.30&   0.00&    1.9&   10.8\\
 1264& 1.2730&  1&  1.975&  2.000&  0.072&   0.43&   0.29&   0.28&    1.6&   14.8\\
 1276& 1.0100&  0&  \nodata&  \nodata&  \nodata&  \nodata&  \nodata&  \nodata&  \nodata&   10.5\\
 1352& 1.3174&  0&  \nodata&  \nodata&  \nodata&  \nodata&  \nodata&  \nodata&  \nodata&    2.6\\
 1362& 1.2729&  1&  2.394&  2.418&  0.049&   0.29&   0.48&   0.23&    2.3&   18.1\\
 1517& 1.1429&  0&  \nodata&  \nodata&  \nodata&  \nodata&  \nodata&  \nodata&  \nodata&   17.1\\
 1533& 1.2779&  1&  \nodata&  \nodata&  \nodata&  \nodata&  \nodata&  \nodata&  \nodata&    4.4\\
 1644& 1.2729&  1&  1.994&  2.018&  0.147&   0.46&   0.54&   0.00&    2.4&    9.5\\
 1698& 1.1419&  0&  \nodata&  \nodata&  \nodata&  \nodata&  \nodata&  \nodata&  \nodata&    7.5\\
 1748& 1.2698&  1&  2.262&  2.286&  0.064&   0.11&   0.63&   0.26&    4.0&   28.8\\
 1763& 1.2749&  1&  2.328&  2.352&  0.043&   0.22&   0.47&   0.31&    6.2&   29.5\\
 1809& 1.2686&  1&  \nodata&  \nodata&  \nodata&  \nodata&  \nodata&  \nodata&  \nodata&    9.5\\
 1888& 1.2771&  1&  2.227&  2.251&  0.044&   0.24&   0.45&   0.31&    4.4&   37.7\\
 2015& 1.2676&  1&  2.079&  2.103&  0.101&   0.16&   0.77&   0.07&    2.6&    9.8\\
 2063& 1.2671&  1&  2.360&  2.384&  0.047&   0.41&   0.39&   0.20&    1.8&   24.1\\
 2111& 1.2798&  1&  2.054&  2.078&  0.076&   0.48&   0.52&   0.00&    1.1&   11.6\\
 2138& 1.1363&  0&  \nodata&  \nodata&  \nodata&  \nodata&  \nodata&  \nodata&  \nodata&    8.5\\
 2336& 0.8731&  0&  2.389&  2.412&  0.044&   0.33&   0.49&   0.18&    2.4&   41.9\\
 2342& 1.2265&  0&  \nodata&  \nodata&  \nodata&  \nodata&  \nodata&  \nodata&  \nodata&    3.4\\
 2369& 1.2667&  1&  2.210&  2.234&  0.122&   0.59&   0.41&   0.00&    0.8&    8.5\\
 2417& 1.2636&  1&  2.195&  2.219&  0.132&   1.00&   0.00&   0.00&    1.4&    9.6\\
 2450& 1.1366&  0&  2.351&  2.375&  0.092&   0.00&   0.79&   0.21&    2.8&    8.9\\
 2497& 1.0823&  0&  \nodata&  \nodata&  \nodata&  \nodata&  \nodata&  \nodata&  \nodata&    5.8\\
 2600& 1.2287&  0&  2.219&  2.243&  0.095&   0.19&   0.39&   0.42&    2.1&   13.4\\
 2624& \nodata&  2&  \nodata&  \nodata&  \nodata&  \nodata&  \nodata&  \nodata&  \nodata&    6.9\\
 2651& 1.2278&  0&  2.440&  2.464&  0.101&   0.62&   0.21&   0.16&    2.0&   23.1\\
 2702& 1.2645&  1&  2.558&  2.582&  0.052&   0.23&   0.48&   0.29&    1.4&    9.1\\
 2735& 1.2669&  1&  2.268&  2.292&  0.059&   0.33&   0.57&   0.11&    1.9&   14.4\\
 2772& 1.2706&  1&  2.132&  2.156&  0.209&   0.68&   0.00&   0.32&    3.1&   22.9\\
 2943& 1.2677&  1&  2.169&  2.193&  0.078&   0.41&   0.00&   0.59&    1.6&   17.7\\
 2989& 1.2643&  1&  2.044&  2.068&  0.089&   0.09&   0.69&   0.22&    3.0&   24.2\\
 3030& 1.2272&  0&  2.268&  2.292&  0.057&   0.51&   0.49&   0.00&    2.0&   14.5\\
 3074& 1.1407&  0&  2.180&  2.203&  0.081&   0.25&   0.42&   0.33&    2.2&   21.6\\
 3090& 1.2780&  1&  2.320&  2.344&  0.040&   0.32&   0.45&   0.23&    7.8&   46.6\\
 3144& 1.3192&  0&  \nodata&  \nodata&  \nodata&  \nodata&  \nodata&  \nodata&  \nodata&   13.3\\
 3384& 1.1954&  0&  \nodata&  \nodata&  \nodata&  \nodata&  \nodata&  \nodata&  \nodata&    7.7\\
 3426& 1.2756&  1&  \nodata&  \nodata&  \nodata&  \nodata&  \nodata&  \nodata&  \nodata&    6.2\\
 3461& 0.6546&  0&  \nodata&  \nodata&  \nodata&  \nodata&  \nodata&  \nodata&  \nodata&   12.0\\
 3505& 0.5700&  0&  \nodata&  \nodata&  \nodata&  \nodata&  \nodata&  \nodata&  \nodata&   30.0\\
 9001& 1.0745&  0&  \nodata&  \nodata&  \nodata&  \nodata&  \nodata&  \nodata&  \nodata&    4.1\\
\enddata
\tablenotetext{a}{Adopted membership: 1 -- galaxy is a member of RXJ0848.6+4453; 
0 -- galaxy is not a member of RXJ0848.6+4453; 2 -- redshift cannot be determined.}
\tablenotetext{b}{Velocity dispersions corrected to a standard size aperture equivalent to a
circular aperture with diameter of 3.4 arcsec at the distance of the Coma cluster.}
\tablenotetext{c}{S/N per {\AA}ngstrom in the rest frame of the galaxy. The wavelength 
interval was chosen based on the redshift of the galaxy as follows: 
redshift 1.00-1.35 -- 3750-4100 {\AA}; 
redshift 0.60-1.00 -- 4100-4600 {\AA};
redshift <0.60 -- 4600-5200 {\AA}.
For ID 336 and ID 2624 a redshift of 1.27 was assumed for the S/N calculation.}
\end{deluxetable*}

\begin{deluxetable*}{r rrr rrr rrr rrr}
\tablecaption{RXJ0848.6+4453: Line Indices and [\ion{O}{2}] for Cluster Members\label{tab-RXJ0848line} }
\tabletypesize{\scriptsize}
\tablewidth{0pc}
\tablehead{
\colhead{ID} & \colhead{CN3883}& \colhead{$\sigma _{\rm CN3883}$} & \colhead{CaHK}& \colhead{$\sigma _{\rm CaHK}$} 
& \colhead{D4000\tablenotemark{a}} & \colhead{$\sigma _{\rm D4000}$}
& \colhead{H$\zeta _{\rm A}$}& \colhead{$\sigma _{\rm H\zeta _A}$} 
& \colhead{EW [\ion{O}{2}]} & \colhead{$\sigma _{\rm EW[OII]}$}
& \colhead{flux([\ion{O}{2}])\tablenotemark{b}} & \colhead{$\sigma _{\rm flux([OII]}$} 
} 
\startdata
\enddata
 240& 0.097& 0.013& 26.60& 0.65& 2.169& 0.015& 7.85& 0.24& 5.0& 2.0& 1.23& 0.19\\
 654& \nodata& \nodata& \nodata& \nodata& \nodata& \nodata& \nodata& \nodata& 183.5& 102.9& 4.10& 0.40\\
 807& \nodata& \nodata& \nodata& \nodata& \nodata& \nodata& \nodata& \nodata& 44.1& 2.5& 6.43& 0.15\\
 1123& \nodata& \nodata& \nodata& \nodata& \nodata& \nodata& \nodata& \nodata& 155.7& 123.5& 5.97& 0.32\\
 1264& \nodata& \nodata& \nodata& \nodata& 2.139& 0.019& \nodata& \nodata& 2.9& 0.6& 0.57& 0.11\\
 1362& 0.179& 0.013& 18.37& 0.73& 1.962& 0.014& 3.32& 0.34& 15.6& 2.7& 3.57& 0.14\\
 1533& \nodata& \nodata& \nodata& \nodata& \nodata& \nodata& \nodata& \nodata& 25.9& 8.3& 2.93& 0.17\\
 1644& \nodata& \nodata& 14.00& 1.59& 2.115& 0.030& 12.68& 0.37& 9.9& 2.6& 1.77& 0.20\\
 1748& 0.122& 0.009& 22.59& 0.50& 2.291& 0.011& 2.90& 0.25& \nodata& \nodata& \nodata& \nodata\\
 1763& 0.202& 0.008& 22.41& 0.40& 2.366& 0.010& 3.06& 0.23& 17.0& 0.5& 8.36& 0.16\\
 1809& \nodata& \nodata& \nodata& \nodata& \nodata& \nodata& \nodata& \nodata& 24.2& 9.9& 2.10& 0.14\\
 1888& 0.158& 0.007& 19.19& 0.34& 2.164& 0.007& 2.09& 0.20& 9.1& 0.3& 5.36& 0.17\\
 2015& \nodata& \nodata& \nodata& \nodata& 3.154& 0.042& \nodata& \nodata& \nodata& \nodata& \nodata& \nodata\\
 2063& 0.070& 0.010& 13.85& 0.62& 1.893& 0.010& 3.65& 0.25& 14.6& 2.2& 4.63& 0.14\\
 2111& 0.113& 0.021& 11.41& 1.24& 1.704& 0.019& 4.62& 0.51& 8.0& 1.0& 1.81& 0.15\\
 2369& 0.099& 0.030& 20.23& 1.63& 2.103& 0.033& 6.64& 0.58& 29.6& 8.4& 1.88& 0.16\\
 2417& 0.116& 0.023& \nodata& \nodata& 1.544& 0.021& 4.69& 0.48& 40.8& 6.5& 7.29& 0.30\\
 2702& 0.156& 0.026& 27.68& 1.49& 2.136& 0.031& 4.29& 0.62& \nodata& \nodata& \nodata& \nodata\\
 2735& 0.089& 0.016& 14.51& 1.04& 1.853& 0.017& 1.99& 0.40& 3.5& 0.9& 0.91& 0.11\\
 2772& \nodata& \nodata& \nodata& \nodata& 1.568& 0.009& \nodata& \nodata& 32.3& 1.4& 10.29& 0.15\\
 2943& 0.188& 0.014& 19.82& 0.77& 2.081& 0.015& 2.60& 0.39& \nodata& \nodata& \nodata& \nodata\\
 2989& \nodata& \nodata& \nodata& \nodata& 2.165& 0.012& \nodata& \nodata& 1.9& 0.3& 0.67& 0.11\\
 3090& 0.174& 0.005& 19.99& 0.27& 2.065& 0.006& 4.17& 0.14& \nodata& \nodata& \nodata& \nodata\\
\tablenotetext{a}{Measured as $\rm D4000_{short}$, cf.\ Eq.\ \ref{eq-D4000}.}
\tablenotetext{b}{Relative flux in $10^{-15}\,{\rm ergs\,cm^{-2} sec^{-1}}${\AA}$^{-1}$
measured within the spectral aperture.}
\tablecomments{The indices have been corrected for galaxy velocity dispersion and aperture corrected.}
\end{deluxetable*}

\clearpage

\begin{figure*}
\epsfxsize 17.5cm
\epsfbox{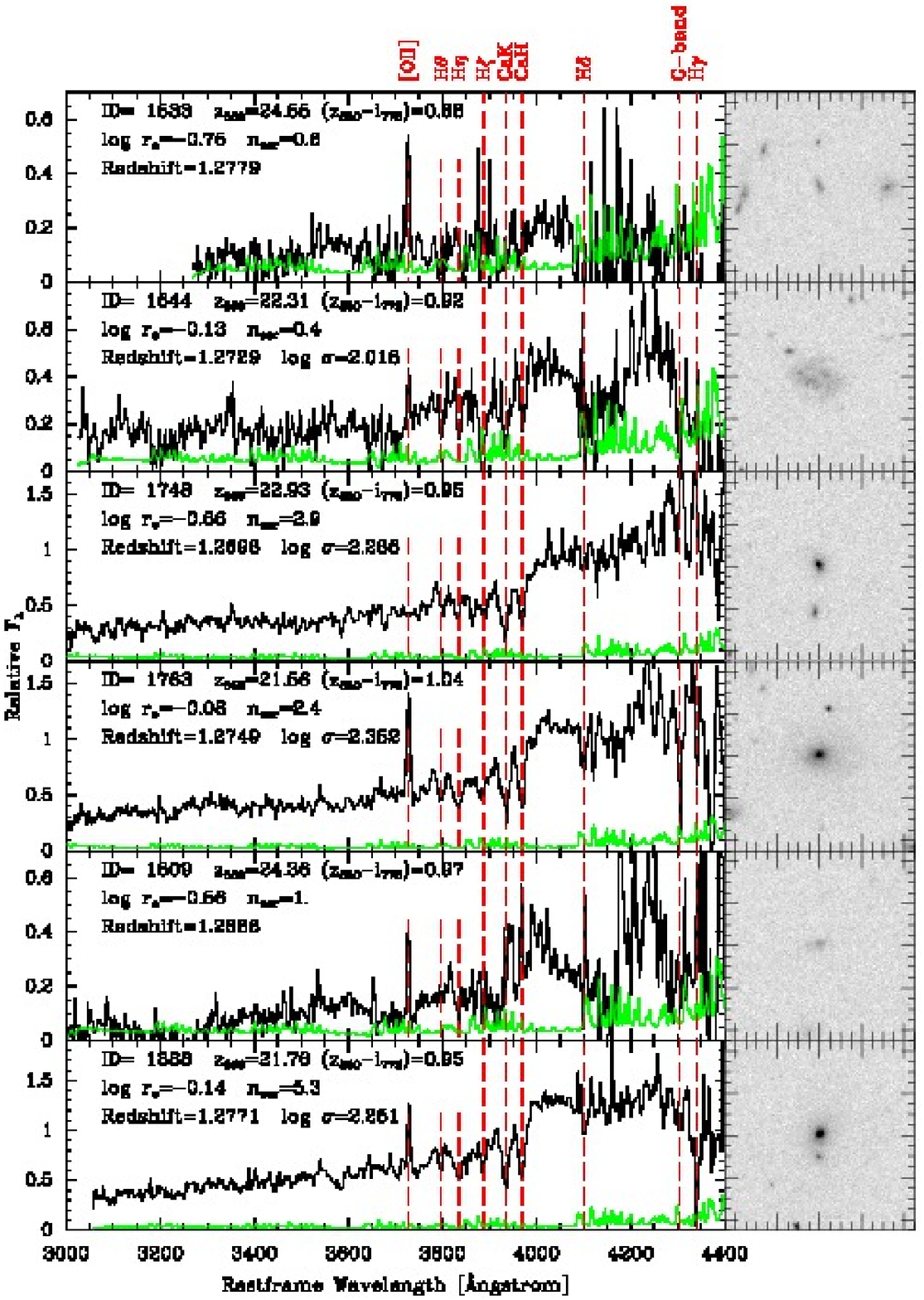}
\vspace{-1cm}
\center{Fig.\ \ref{fig-specRXJ0848} -- {\em Continued.}}
\end{figure*}
\begin{figure*}
\epsfxsize 17.5cm
\epsfbox{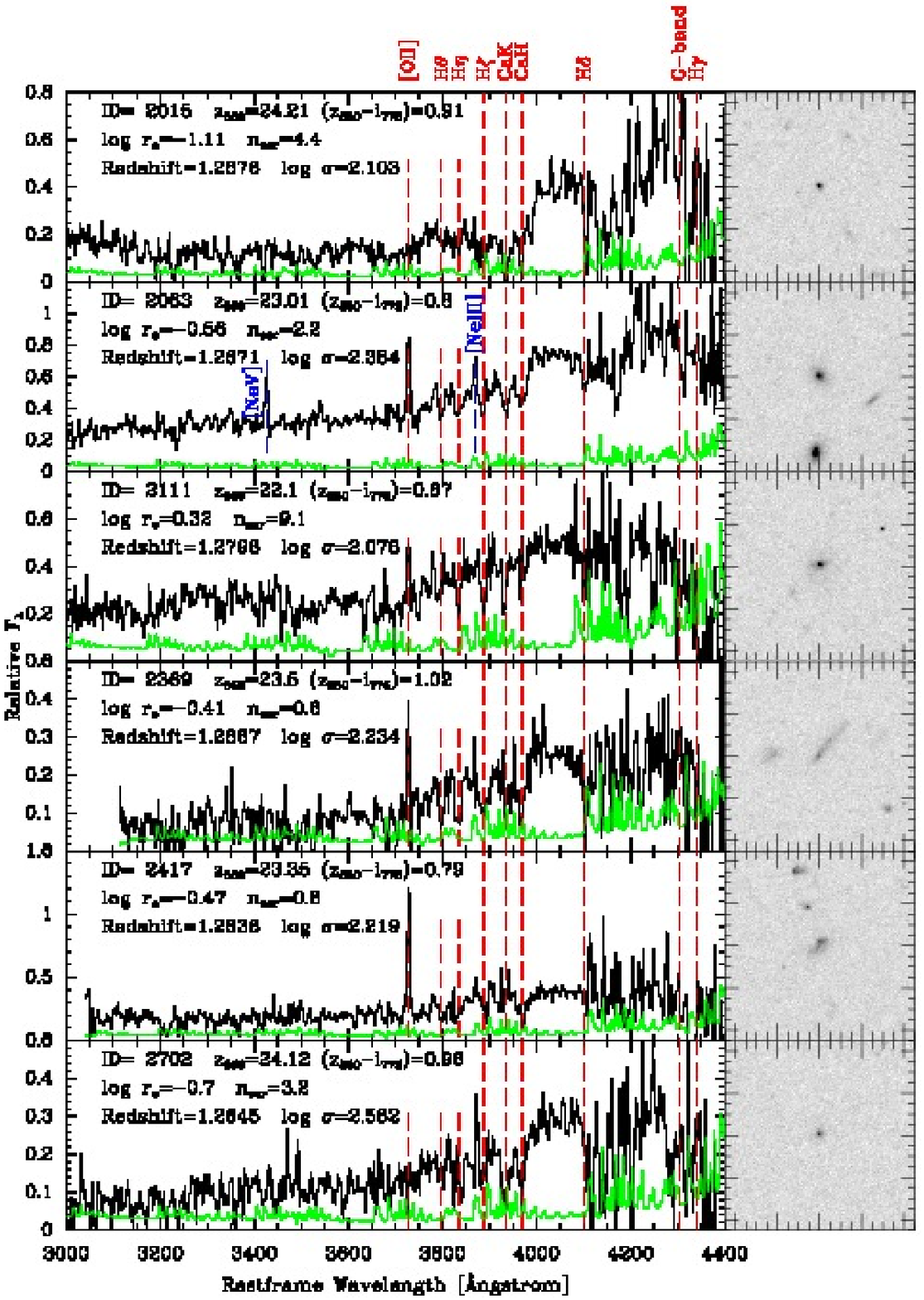}
\vspace{-1cm}
\center{Fig.\ \ref{fig-specRXJ0848} -- {\em Continued.}}
\end{figure*}
\begin{figure*}
\epsfxsize 17.5cm
\epsfbox{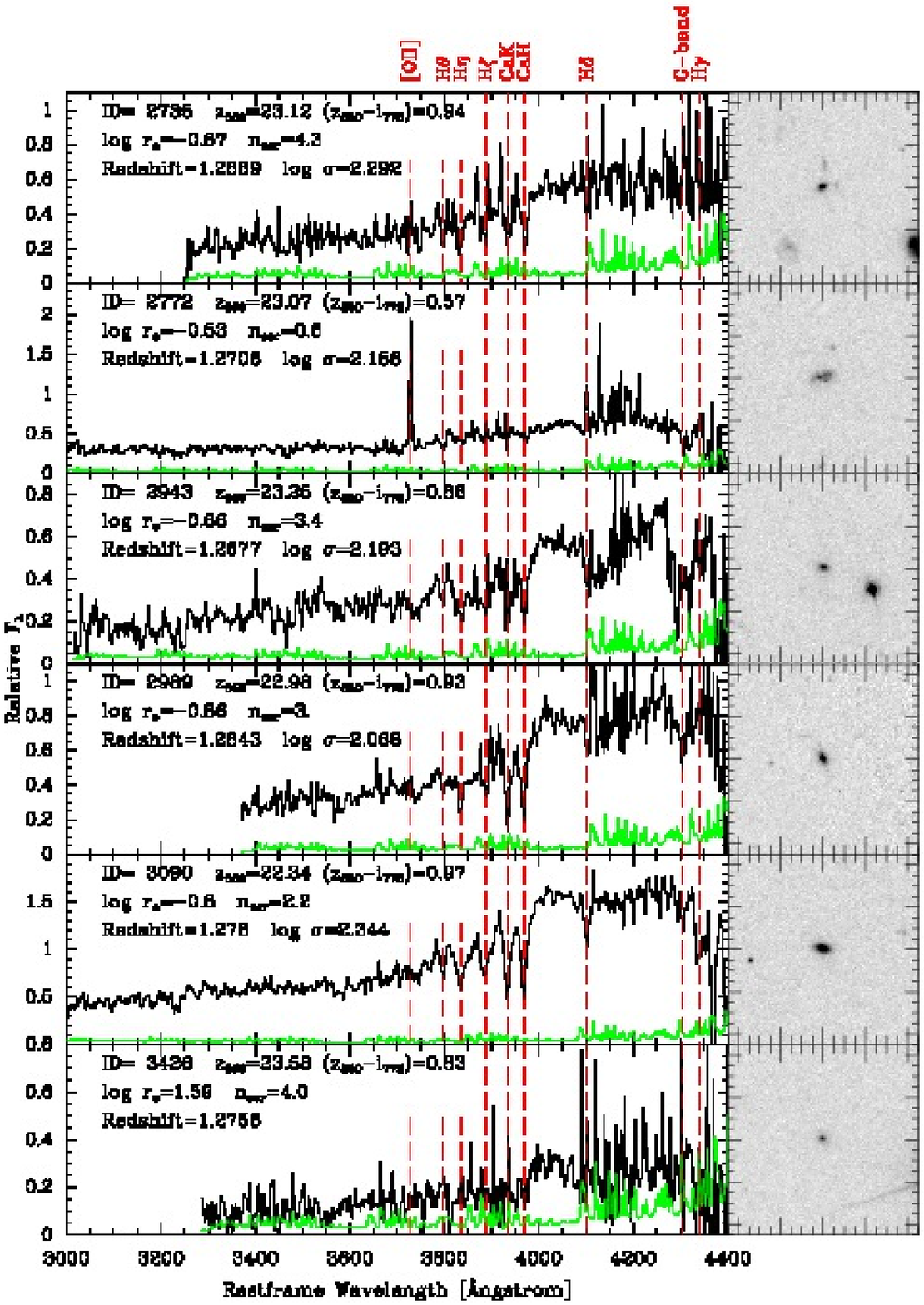}
\vspace{-1cm}
\center{Fig.\ \ref{fig-specRXJ0848} -- {\em Continued.}}
\end{figure*}

\end{document}